\documentclass[twocolumn, tighten]{aastex63}

\newcommand{\Rmnum}[1]{\expandafter\@slowromancap\romannumeral #1@}

\usepackage{amsmath,amsthm,amssymb}
\usepackage{multirow}
\usepackage{hyperref}

\begin{document}

\title{Extended X-ray emission associated with the radio lobes and the environments of 60 radio galaxies}

\author{Ajay Gill}
\affiliation{David A. Dunlap Dept. of Astronomy and Astrophysics, University of Toronto, 50 St. George Street, Toronto, ON, Canada M5S 3H4}
\affiliation{Dunlap Institute for Astronomy and Astrophysics, University of Toronto, 50 St. George Street, Toronto, ON, Canada M5S 3H4}
\author{Michelle M. Boyce}
\affiliation{Department of Physics and Astronomy, University of Manitoba, Winnipeg, MB R3T 2N2, Canada}
\author{Christopher P. O'Dea}
\affiliation{Department of Physics and Astronomy, University of Manitoba, Winnipeg, MB R3T 2N2, Canada}
\author{Stefi A. Baum}
\affiliation{Faculty of Science, University of Manitoba, Winnipeg, MB R3T 2N2, Canada}
\author{Preeti Kharb}
\affiliation{National Centre for Radio Astrophysics, TIFR, Post Bag 3, Ganeshkhind, Pune 411007, India}
\author{Neil Campbell}
\affiliation{Physics Department, 2320 Chamberlin Hall, University of Wisconsin-Madison,1150 University Avenue, Madison, Wisconsin 53706, USA}
\author{Grant R. Tremblay}
\affiliation{Harvard-Smithsonian Center for Astrophysics, 60 Garden St., Cambridge, MA 02138, USA}
\author{Suman Kundu}
\affiliation{Centre for Nano and Soft Matter Sciences, P.B.No.1329, Prof UR Rao Road, Jalahalli, Bengaluru, 560013, India}

\begin{abstract}
This paper studied the faint, diffuse extended X-ray emission associated with the radio lobes and the hot gas in the intracluster medium (ICM) environment for a sample of radio galaxies. We used shallow ($\sim 10$ ks) archival \textit{Chandra} observations for 60 radio galaxies (7 FR I and 53 FR II) with $0.0222 \le z \le 1.785$ selected from the 298 extragalactic radio sources identified in the 3CR catalog. We used Bayesian statistics to look for any asymmetry in the extended X-ray emission between regions that contain the radio lobes and regions that contain the hot gas in the ICM. In the \textit{Chandra} broad band ($0.5 - 7.0$ keV), which has the highest detected X-ray flux and the highest signal-to-noise ratio, we found that the non-thermal X-ray emission from the radio lobes dominates the thermal X-ray emission from the environment for $\sim 77\%$ of the sources in our sample. We also found that the relative amount of on-jet axis non-thermal emission from the radio lobes tends to increase with redshift compared to the off-jet axis thermal emission from the environment. This suggests that the dominant X-ray mechanism for the non-thermal X-ray emission in the radio lobes is due to the inverse Compton upscattering of cosmic microwave background (CMB) seed photons by relativistic electrons in the radio lobes, a process for which the observed flux is roughly redshift independent due to the increasing CMB energy density with increasing redshift. 
\end{abstract}

 
\section{Introduction}  \label{sec:intro}

Radio galaxies provide a number of important challenges to astrophysics, including those of jet launching, propagation, and effects on the host galaxy and environment  \citep[see e.g.][]{Heckman_2014, Padovani_2016, Tadhunter_2016, Padovani_2017,Blandford2019,Hardcastle2020}. The dominant emission mechanism in the radio wavelengths is synchrotron emission, where radiation is emitted by relativistic electrons gyrating through magnetic fields. Synchrotron emission is non-thermal in origin and is typically parameterized by a power law of the form $S_{\nu} \propto \nu^{-\alpha}$, where $S_{\nu}$ is the flux density, 
$\nu$ is the frequency, and $\alpha$ is the spectral index.  

\citet{Fanaroff} suggested that radio galaxies could be categorized into two general luminosity/morphology classes, namely FR I and FR II. For FR I (FR II) sources, the ratio of the distance between the brightest spots of radio emission on either side of the center to the full extent of the source is less (greater) than 0.5. The FR II sources tend to have higher radio powers  ($>$ 10$^{25}$ W Hz$^{-1}$) than the FR Is.  The FR types also separate in the radio luminosity versus optical host galaxy luminosity plane (see e.g. \citealp{Ledlow_1996, buttiglione}). Detailed discussions of the differences between FR I and FR II are given by e.g., \citet{Bridle1984} and \citet{Baum1995}.

X-ray observations provide for a near complete detection of {active galactic nuclei} (AGNs) with minimal contamination from non-AGN sources \citep{Padovani_2017}. The reasons for this are: (i) X-ray emission from AGN seems to be nearly universal; (ii) X-rays can penetrate through high column densities of gas and dust; {and} (iii) X-ray emission from host galaxy astrophysical processes are typically weak compared to the AGN \citep{Brandt_2015}. Radio galaxies consist of a compact core at the nucleus, extended radio lobes, hotspots, and collimated radio jets, all of which may radiate X-ray emission (see e.g. \citealp{arXiv:1609.07145, Mingo, deVries, massaro3Csnapshot}).

Several emission mechanisms have been proposed for the X-ray emission from radio galaxies \citep[see e.g.][]{harris2002,harris2006}.
Thermal bremsstrahlung emission  does not provide a satisfactory model for the X-ray emission associated with  radio lobes and hotspots because (i) the required amount of hot gas is large ($\sim 10^{10}\,\, M_{\odot}$), (ii) the gas needs to be far from the host galaxy, and (iii) the predicted Faraday rotation and depolarization are not observed \citep{harris2002}.
The bright compact hotspots at the end of jets can be consistent with either synchrotron or synchrotron self Compton (SSC) emission \citep{Perley,Hardcastle2004,Mingo}. 
 In the diffuse radio lobes, the dominant photon field is the Cosmic Microwave Background (CMB) \citep[e.g.,][]{croston2005}, and inverse Compton (IC) scattering of the CMB photons by relativistic electrons {has been found to be} consistent with X-ray observations of the radio lobes
 \citep[see e.g.][]{Tavechhio2000,Celotti,harris2002,croston2005,Kharb12b,Stanley15,ineson_2017,Mingo}. 
 

For the low-luminosity FR I radio sources, the dominant extended emission mechanism in X-ray, optical, and radio from the jets is thought to be due to the synchrotron process \citep[e.g.,][]{Sambruna_2004,Worrall09,Kharb12a}. This is supported by the intrinsic variability found in the knots of the M87 jet \citep{Harris_2006}, and the fact that in most cases the X-ray spectral index $\alpha_{x}$ is greater than 1 and much greater than the radio spectral index $\alpha_{r}$, as well as the relative morphologies in the X-ray, optical, and radio.


{The purpose of this study was to search for faint, diffuse extended X-ray emission associated with the radio lobes and the hot gas in the intracluster medium (ICM) environments for a sample of 60 radio galaxies observed in shallow ($\sim 10$ ks) archival \textit{Chandra} observations. Since the exposure times for the \textit{Chandra} observations of the sources in our sample were short, detailed spectral modeling of the extended emission from the radio lobes and the environments was not possible. Instead, we studied the properties of the thermal and non-thermal extended emission of the sources in our sample by considering the spatial distribution of the emission.} We did this by comparing the X-ray emission  in regions that contain the radio lobes and regions that do not contain the radio lobes {in the hot ICM environments}. Our technique (described in \textsection\ref{ap}) allows us to search for faint, diffuse X-ray emission. 
We might expect to see asymmetry in the diffuse X-ray emission, since as discussed above, there has been previous detection of non-thermal emission  (synchrotron, SSC, or IC/CMB)  associated with the radio plasma. 
Furthermore, the radio source itself can also influence the ambient thermal gas in the intergalactic and intracluster medium, forming bubbles/cavities in the ICM \citep[see e.g.][]{rafferty2006,McNamara:2007ww,cavagnolo2010,sullivan2011,hlavacek2011}.

This paper is organized as follows. In \textsection\ref{sec:sample}, we present the sample selection criteria. {The overview of the X-ray and radio data used in this study is presented in \textsection\ref{sec:data_information}}. The archival data analysis is described in \textsection\ref{ap}, consisting of the data reduction procedure (\textsection\ref{datared}), {the quadrant selection and} the point source blanking procedure (\textsection\ref{blanking}), the sky background estimation (\textsection\ref{bkg}), and the calculation of the X-ray emission asymmetry (\textsection\ref{sec:asymmetric}). {The results and the discussion are presented in \textsection\ref{sec:results} and \textsection\ref{sec:discussion}, respectively}. During this study, we found some sources with X-ray emission spatially correlated with the radio hotspots. These sources were not included in our sample and are instead
{reported} in \textsection\ref{sec:report}. In Appendix \textsection{}\ref{ap:bayesian}, we present the analytical framework for the Bayesian analysis used to compute the ``ON/OFF jet-axis asymmetry'' parameter, $R$ (see Equation \ref{eq:r} in \textsection{}\ref{sec:asymmetric}). In Appendix \textsection{}\ref{ap:r_values}, we present the $R$ values for each source by band. In Appendix \textsection{}\ref{ap:science_circles}, we present the properties of the science aperture as well as the X-ray and radio data for each source. Throughout this paper, we assume ${H}_{\mathrm{0}}$ = 72 km s$^{-1}$ Mpc$^{-1}$, $\Omega_{\mathrm{M}}$ = 0.27, $\Omega_{\Lambda}$ = 0.73, consistent with \citet{arXiv:0803.0586}. 

\section{Sample selection criteria}
\label{sec:sample}

{The Third Cambridge catalog (3C) of 471 bright radio sources observed at 159 MHz was first published in 1959 \citep{edge}. The 3C catalog was revised (3CR) in 1962 with observations at 178 MHz \citep{bennett}. The 3CR catalog has a flux limit of 9 Jy at 178 MHz and covers the entire northern hemisphere north of $-5^{\circ}$ in declination and contains 328 radio sources. In 1985, \citet{Spinrad_1985} provided an update to the 3CR catalog with revised positions, redshifts, magnitudes, and identifications, and presented a list of 298 extragalactic radio sources, which forms the initial sample for this study. These 298 extragalactic sources form a heterogeneous sample in redshift and source  type.}

\begin{figure}[htb!]
    \centering
    \includegraphics[width=25em]{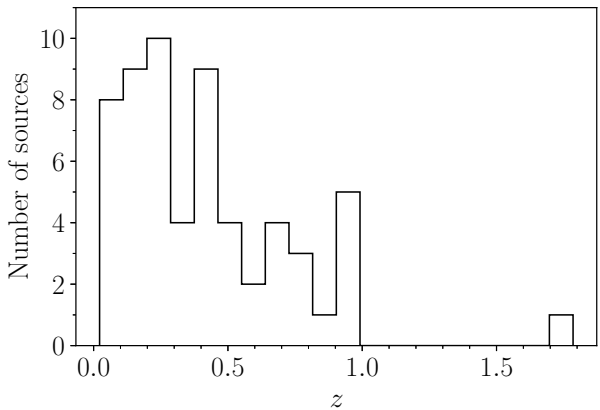}
    \caption{{Redshift distribution of the sample in this study, consisting of 60 extragalactic radio sources (7 FR I and 53 FR II) with redshifts between $0.0222 \le z \le 1.785$ selected from the 298 extragalactic radio sources from the 3CR catalog \citep{bennett}.}}
    \label{fig:z_dist}
\end{figure}

{
We selected 3CR sources which have \textit{Chandra} X-ray data and  have Very Large Array (VLA) radio data that shows the radio lobes and have a ``science" aperture (see \textsection \ref{blanking}) that does not exceed the boundaries of the \textit{Chandra} ACIS chip. We excluded three sources whose radio structure extended beyond the defined on-axis quadrants (see \textsection \ref{blanking}). 
Since the purpose of this study was to search for faint, diffuse extended X-ray emission associated with the radio lobes and the hot gas in the environment, we excluded sources with bright previously detected features  as determined by visual inspection of the sources for bright X-ray jets or hotspots, as well as by using the XJET\footnote{\texttt{https://hea-www.harvard.edu/XJET/}} catalog as a guide.   Our final sample consists of 60 sources (7 FR I and 53 FR II). The redshift distribution of the sample is shown in Figure \ref{fig:z_dist}, and is between $0.0222 \le z \le 1.785$.}

\section{X-ray and radio data} \label{sec:data_information}
We used archival \textit{Chandra} \citep{Weisskopf_2002} observations for the radio sources in our sample. The low background levels, high angular resolution, and the collecting area of \textit{Chandra} allows it to detect sources three orders of magnitude fainter than previous generation X-ray observatories. The radio images for the sources in our sample consisted of images observed with the VLA retrieved from the NRAO Science Data Archive\footnote{\texttt{https://archive.nrao.edu/archive/}} as well as from colleagues. $L$-band ($1.0 - 2.0$ GHz) radio images were preferentially used due to their superior detection of the radio lobes. Higher frequency $C$-band ($4.0 - 8.0$ GHz) and $X$-band ($8.0 - 12.0$ GHz) images were used when $L$-band images for a particular source were unavailable. {The archival \textit{Chandra} data is mostly from the snapshot observations of Massaro and collaborators \citep{Massaro12,Massarobetween,Massaro_survey,massaro3Csnapshot} with exposure times of $\sim$ 10 ks. The FR classifications of the sources in our final sample were taken from \citet{arXiv:1609.07145}}.



\section{Archival Data Analysis} \label{ap}
\subsection{Data Reduction Procedure} \label{datared}
The X-ray event files of the 
sources {in our sample} were taken from the $Chandra$ ChaSeR data archive.\footnote{\texttt{http://cda.harvard.edu/chaser/} \label{footnote1}} 
{The $Chandra$ datasets are listed in Table \ref{tab:raw_quadrants}.}

A circular science aperture consisting of four quadrants of equal area was overlaid on the source to determine the extended X-ray emission asymmetry between regions containing the radio lobes and those that do not. We considered four different X-ray energy bands defined by the $Chandra$ ACIS Science Energy Bands\footnote{\texttt{http://cxc.harvard.edu/csc/columns/ebands.html} \label{footnote3}} with effective energies: soft ($0.5 - 1.2$ keV), medium ($1.2 - 2.0$ keV), hard ($2.0 - 7.0$ keV), and broad ($0.5 - 7.0$ keV).

The $Chandra$ data were reduced following the standard procedure described in the \emph{Chandra Interactive Analysis of Observations\/} (CIAO) threads,\footnote{\texttt{http://cxc.harvard.edu/ciao/guides/index.html}} using CIAO version 4.8 and \emph{Chandra Calibration Database\/} version CALDB version 4.7.2. The  \texttt{chandra\textunderscore repro} command was used to process the data from the standard data distribution. The parameters \texttt{destreak},  \texttt{badpixel}, and \texttt{process\textunderscore events} were used to detect streak events and create a bad pixel and level 2 event file. 

For sources with multiple observations, \texttt{reproject\textunderscore obs} was used to reproject the observations to a common tangent point. The native ACIS pixel size is 0.492$^{\prime\prime}$. A binning factor of 0.5 corresponding to 0.246$^{\prime\prime}$ per pixel was used to adequately sample the point spread function. Four exposure-corrected images were created using the \texttt{flux\textunderscore obs} command for the soft, medium, hard, and broad bands, where the intensity level in each pixel was in units of counts s$^{-1}$. 

\subsection {{Quadrant selection and point source blanking procedure}} \label{blanking} 
\begin{figure} [htb!]
\centering	
\includegraphics[width=22em]{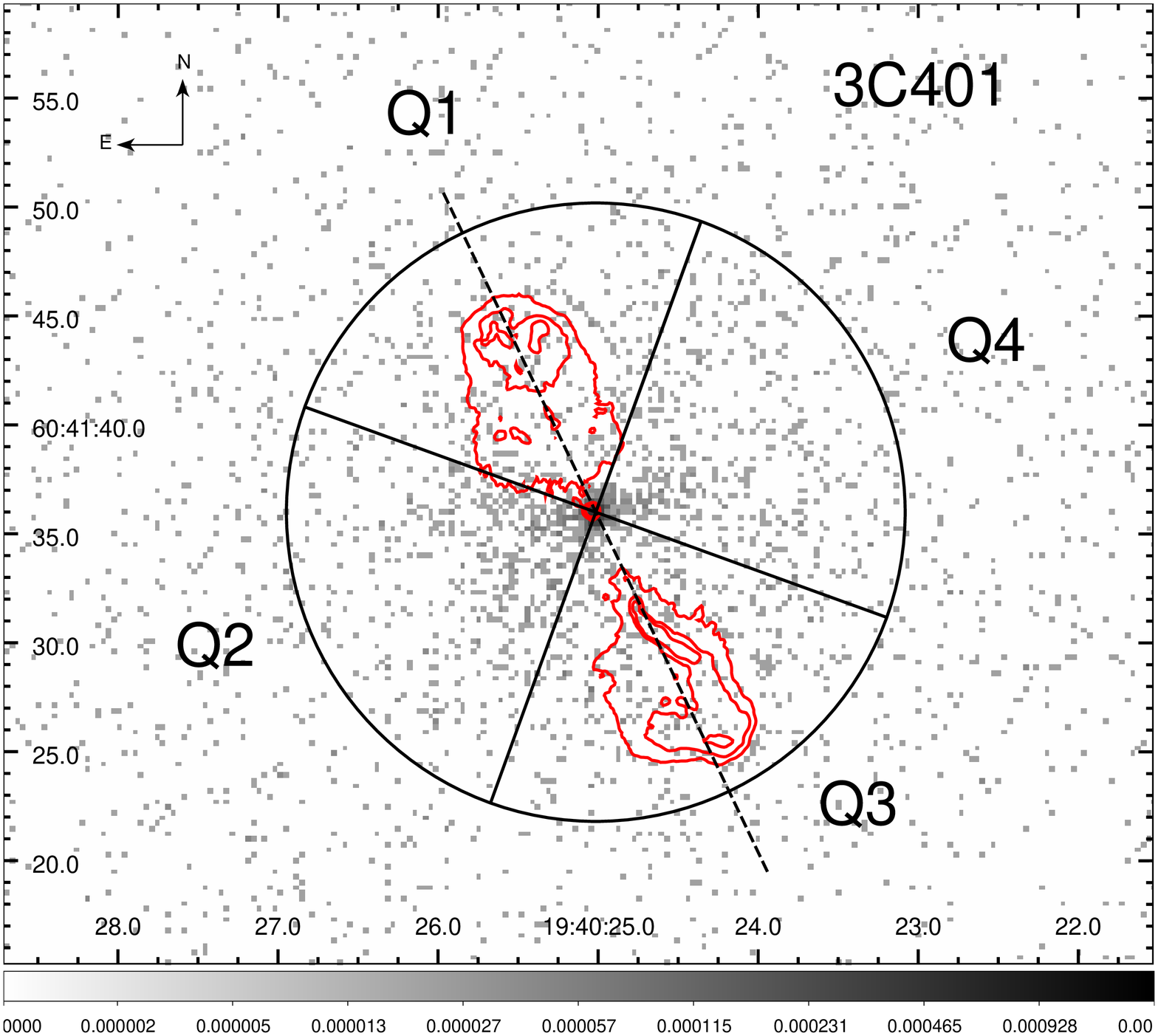}	
\includegraphics[width=22em]{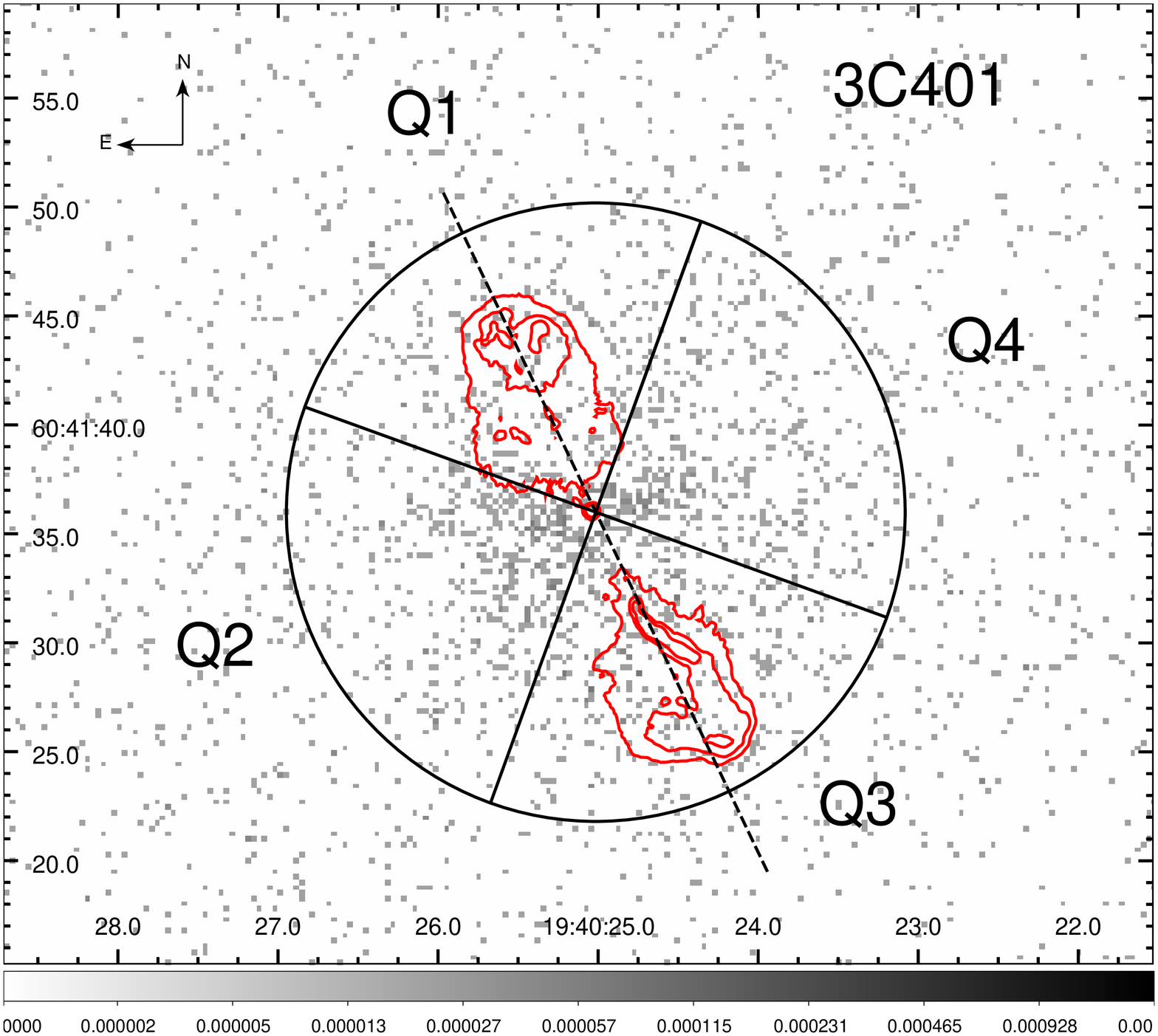}
\caption{3C 401 X-ray broadband ($0.5 - 7.0$ keV) image before (top) and after (bottom) blanking. X-ray emission: gray pixels. Radio emission: red contours. The X-ray data consists of a combination of two observations (with \textit{Chandra} IDs 3083 and 4370) with exposure times of 22.66 ks and 24.85 ks, respectively. The radio image is an $X$-band VLA image.}
\label{fig:example}
\end{figure}

For each source in our sample, the radio image was used as a reference to construct a circular ``science" aperture  with four quadrants of equal area. Two quadrants were on-axis (each containing a radio lobe), and the other two quadrants were off-axis (each not containing a radio lobe). The radius of the aperture was manually chosen by visual inspection for each individual source by careful consideration of the extent of the radio lobes seen in the selected radio images, such that both radio lobes are sufficiently contained within the aperture (see Figure \ref{fig:example}). 
{The science apertures are also given in Table \ref{tab:raw_quadrants} and the
counts in the apertures are given in Table \ref{tb:on_off_b_counts}. Three sources (3C 76.1, 3C 288, and 3C 433) which had radio structure that extended beyond the nominal on-axis quadrants were excluded from the analysis.} The science aperture was then overlaid onto the exposure-corrected X-ray image of the source. 
The broadband $Chandra$ image was used for selecting the center {of the science aperture}, since it contains the widest energy range of all bands and hence the highest signal-to-noise ratio. 

{We note that about 97\% of 3CR radio nuclei are detected in short observations by $Chandra$
\citep{Massaro12,Massarobetween,Massaro_survey,massaro3Csnapshot}. Thus, it was relatively simple to overlay the science aperture (determined from the radio image) onto the $Chandra$ image and center it on the X-ray nucleus.  Any miscentering of a few pixels between the radio and X-ray core should not affect the on-jet and off-jet axis counts, since (i) the X-ray nucleus region is blanked, and (ii)  the extent and area of the radio lobes and hence the quadrants is much larger than the X-ray nucleus. Therefore, a more sophisticated centroiding algorithm was not necessary for this study.} 
 
After the science aperture was constructed, the pixels of the point source at the nucleus were replaced with new values that were the mean of the Poisson distribution of pixel values in the background diffuse emission. This was done to remove unwanted nuclear emission from the quadrants, since the aim of the study was to investigate the asymmetry in only the diffuse extended X-ray emission between the on and off-axis radio lobe quadrants.

The \texttt{dmfilth} tool was used to fill the nuclear pixels with the background diffuse emission pixels, after which the \texttt{dmimgcalc} tool was used to obtain an exposure-corrected X-ray image, which we call the ``blanked'' image. {The radius of the exclusion circle used for blanking the point source at the nucleus was manually selected by visual inspection of (i) the overdensity of bright pixels associated with the X-ray nucleus in the broadband image, and (ii) comparison to the radio core. This radius varied for each source, and was typically between 2.5 to 5 arcseconds. For the \texttt{dmfilth} tool, the background region was an annulus around the exclusion region. The radius of the background region was chosen to be three times larger than the radius of the exclusion region, with a gap between the exclusion region and background region of 0.5 pixels to ensure there is sufficient separation between the inner radius of the background region and the exclusion region.} 

{To test whether choosing a larger background radius has any affect on the ON$-$OFF axis counts, we looked at the source 3C 401 as a test source and used a background radius that is ten times larger than the exclusion radius. We found no difference in the ON$-$OFF axis ratio between the cases whether the radius is three times larger or ten times larger than the exclusion region. Therefore, a background region with a three times larger radius than the exclusion region should be sufficient for the purpose of blanking the nucleus in this study.} The blanking procedure we followed is described in further detail in \emph{An Image of Diffuse Emission} in the \emph{CIAO 4.9 Science Threads}.\footnote{\texttt{http://cxc.harvard.edu/ciao/threads/diffuse$\_$emission/}} Figure \ref{fig:example} shows an example of the quadrant selection and point source blanking procedure with the unblanked and blanked image of 3C 401. 


\subsection{Background Estimation} \label{bkg}

After the X-ray data were reduced and the point source blanking was complete, the image was also background subtracted. The background level was estimated by constructing between one and three circular ``background" apertures with the same radius as the science aperture. The background apertures were placed away from the science aperture in the background regions (containing no point sources), while still being on the same ACIS chip as the science aperture. The background counts were estimated by summing the total number of counts in each background aperture and then taking the median of the three sums from the three apertures. We denote these median background counts for an aperture the size of the science aperture as $B$. We used the median of up to three apertures to estimate the background so that the background estimate is less biased toward the potential variability of the efficiency of the pixels across the ACIS chip. Three background apertures were preferentially used whenever possible. For a few sources, the science aperture was sufficiently large on the ACIS chip that it was only possible to construct only one or two background apertures. {The background counts are given in Table \ref{tb:on_off_b_counts}.}

\subsection{Calculation of the X-Ray Emission Asymmetry}
\label{sec:asymmetric}
The ON$-$OFF jet-axis asymmetry is calculated as

\begin{equation}
    R = \frac{S_{\rm ON}-S_{\rm OFF}}{(S_{\rm ON}+S_{\rm OFF})\,/\,2}\,,\label{eq:r}
\end{equation}

\noindent where 

\begin{equation}
   \left.\begin{array}{l@{\;=\;}c}
         S_{\rm ON} & \text{Counts}_{\rm Q1} + \text{Counts}_{\rm Q3}\\
         S_{\rm OFF} & \text{Counts}_{\rm Q2} + \text{Counts}_{\rm Q4}
    \end{array}\right\}\,
\end{equation} 

\noindent such that $R > 0$ means more on-axis emission, $R = 0$ means symmetric emission, and $R < 0$ means more off-axis emission. The background counts are taken into consideration by 

\begin{equation}
    B_{\rm ON}=B_{\rm OFF}\equiv B/\,r
    \label{eq:bees}
\end{equation}

\noindent where $r = 2$,\footnote{See Appendix~\ref{ap:bayesian}.} and $B$ is the estimated background counts within a background aperture with the same radius as the science aperture on the ACIS chip. As the X-ray data for the sources {in our sample} consist of relatively low counts, background subtraction must be done carefully to avoid negative values. As such, traditional error propagation techniques do not generally apply.

The approach taken here is to model Equation (\ref{eq:r}) using Bayesian analysis, as prescribed by \citet{park2006} and \citet{dyk2001}. The posterior distribution of $R$, given prior information $S_{\rm ON}$, $S_{\rm OFF}$, and $B$, is given by

\begin{eqnarray}
    P(R|S_{\rm ON},S_{\rm OFF},B)&=&\sum_{i=0}^{S_{\rm ON}}\sum_{j=0}^{S_{\rm OFF}}\kappa_{ij}(S_{\rm ON},S_{\rm OFF},B,R)\nonumber\\&&\!\!\!\!\!\!\!\!\!\!\!\!\!\!\!\!\times\frac{\Gamma(i+j+S_{\rm ON}+S_{\rm OFF}+2)}{4^{i+j+S_{\rm ON}+S_{\rm OFF}+2}}\,,\label{eq:pr}
\end{eqnarray}

\noindent where

\begin{eqnarray}
    \kappa_{ij}(S_{\rm ON},S_{\rm OFF},B,R)&=&N(i,S_{\rm ON},B)\,N(j,S_{\rm OFF},B)\nonumber\\&&\!\!\!\!\!\!\!\!\!\!\!\!\!\!\!\!\!\!\!\!\times\left(1+\frac{R}{2}\right)^{i+S_{\rm ON}}\left(1-\frac{R}{2}\right)^{j+S_{\rm OFF}}\,,\nonumber\;\;\;\;\;\;
\end{eqnarray}

\noindent and

\begin{equation}
\begin{array}{l}
    N(\ell,S_x,B)=\\\;\;\;\;\;\;\;\;\frac{\displaystyle\frac{\Gamma(S_x-\ell+2B+1)}{\Gamma(\ell+1)\Gamma(S_x-\ell+1)4^{S_x-\ell+2B+1}}}{\displaystyle \sum_{k=0}^{S_x}\frac{\Gamma(S_x-k+2B+1)\Gamma(k+S_x+1)}{\Gamma(k+1)\Gamma(S_x-k+1)4^{S_x-k+2B+1}2^{k+S_x+1}}}\,,\nonumber
\end{array}
\end{equation}

\noindent with $x$ = ON, OFF, $\ell=i,j,$ and $-2\le R\le 2$, as derived in Appendix~\ref{ap:bayesian}. This formula allows us to simultaneously handle small (Poissonian) and large (Gaussian) statistics.

\begin{figure}[htb!]
\centering	
\includegraphics[width=\columnwidth]{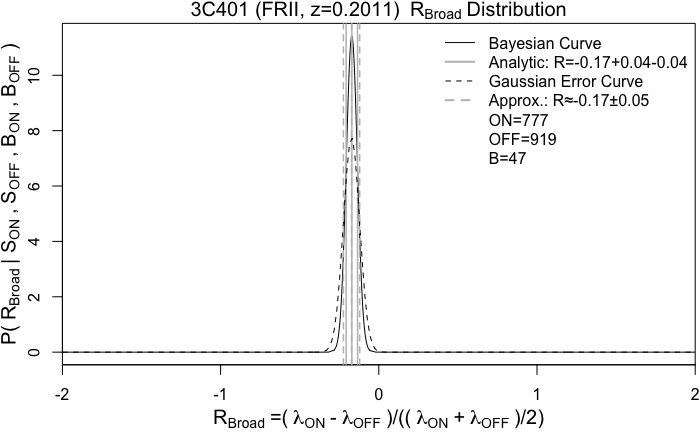}	
\caption{Posterior distribution of $R_{\rm broad}$ for 3C 401 of Figure \ref{fig:example}, with mode $-0.17$ and a 
{68.27}\% Highest Posterior Density (HPD) region \citep{hoff2010} of 
{$[-0.21,\,-0.13]$}, i.e., $R_{\rm broad}=-0.17\pm$
{0.04}. So the X-ray emission is off the jet-axis.}
\label{fig:r_3c401_broad}
\end{figure}

Figure~\ref{fig:r_3c401_broad} shows an example of the $R$ distribution for source 3C 401 in the broad band, depicted in Figure~\ref{fig:example}. Figure \ref{fig:r_3c401_broad} is consistent with the extended X-ray emission being off the jet-axis, i.e., $R_{\rm broad}=-0.17\pm$
{0.04}. Note that this particular example has rather high counts, $S_{\rm ON}=777$ and $S_{\rm OFF}=919$, with relatively low background, $B=47$. The 
{68.27\% ($\sim\pm1\sigma$) Highest Posterior Density (HPD)} region is also symmetrical about its mode. All implying, that Gaussian error propagation should apply: i.e.,

\begin{equation}
    R \simeq \frac{S_{\rm ON}-S_{\rm OFF}}{\displaystyle\frac{1}{2}\left(\!S_{\rm ON}+S_{\rm OFF}-2\frac{B}{r}\!\right)}\pm\sigma_R\label{eq:r_approx}
\end{equation}

\noindent with error,
\begin{equation}
    \sigma_R=\frac{\sqrt{\displaystyle\!\left(\!\!S_{\rm OFF}-\!\frac{B}{r}\!\right)^2\!\!\!\sigma_{S_{\rm ON}}^2+\left(\!\!S_{\rm ON}-\!\frac{B}{r}\!\right)^2\!\!\!\sigma_{S_{\rm OFF}}^2+\frac{\sigma_B^2}{r^2}}}{\displaystyle\frac{1}{4}\left(\!\!S_{\rm ON}+S_{\rm OFF}-2\frac{B}{r}\!\right)^2}\nonumber
\end{equation}

\noindent where
\begin{equation}
    \sigma_N\approx\sqrt{N+0.75}+1\nonumber
\end{equation}

\noindent for counts $N = S_{\rm ON},\, S_{\rm OFF},\, B$  \citep{gehrels1986}. Plugging in the numbers, we find $R_{\rm broad}\simeq-0.17\pm0.05$ which is consistent with the analytical expression, Equation (\ref{eq:pr}). 

\begin{figure}[htb!]
\centering	
\includegraphics[width=\columnwidth]{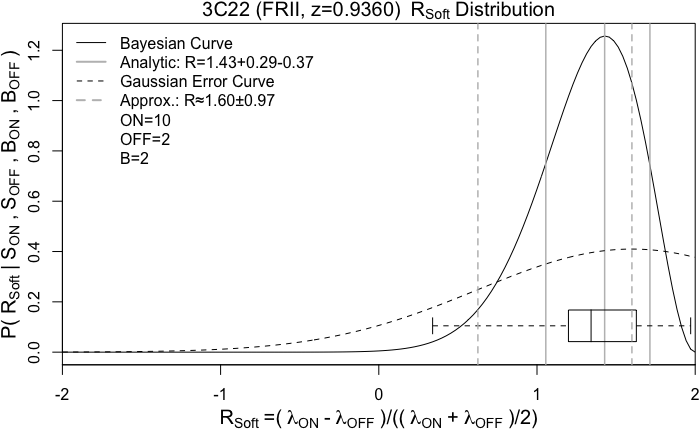}	
\caption{Posterior distribution of $R_{\rm soft}$ for 3C 22, with 
{$R_{\rm soft}=1.43^{+0.29}_{-0.37}$ for the analytic solution, and $R_{\rm soft}\simeq1.60\pm0.97$ for the approximate method. Overlaid on the analytic curve is a box plot, with 99.3\% HPD whiskers, 50\% HPD box, and 1.34 median, providing a 1D method of visualizing a skewed 2D distribution \citep{hoff2010}: see Figure \ref{fig:r_table}}. So the X-ray emission is on the jet-axis.}
\label{fig:r_3c22_soft}
\end{figure}

For low statistics, however, the approximation deviates significantly. Figure \ref{fig:r_3c22_soft} shows an example of the $R$ distribution for source 3C 22, in the soft band, where $S_{\rm ON}=10$, $S_{\rm OFF}=2$, and $B=2$.
Here the analytical expression, Equation (\ref{eq:pr}), yields 
{$R_{\rm soft}=1.43^{+0.29}_{-0.37}$}, whereas the approximation, Equation (\ref{eq:r_approx}), yields $R_{\rm soft} \simeq 1.60\pm0.97$. 
{A more detailed discussion on this subject can be found in \citet{park2006}.} 

{For simplicity, $R\pm1\hat{\sigma}$ shall be taken as shorthand for $R_{-\sigma_1}^{+\sigma_2}$, such that, $R-\sigma_1$ and $R+\sigma_2$ represent the lower and upper 68.27\% HPD ($\sim\pm1\sigma$) bounds, respectively, where $R$ is the mode of the distribution function. This definition is convenient, as it is synonymous for either the analytic or approximate expression, Equations (\ref{eq:pr}) or (\ref{eq:r_approx}), respectively. Unless explicitly stated otherwise, we shall assume $R$ is the mode of the distribution with $\pm1\hat{\sigma}$ confidence intervals.}




\section{Results} \label{sec:results} 

 {Table~\ref{tb:r_stats} (in Appendix \ref{ap:r_values}) shows the $R\pm1\hat{\sigma}$ values obtained from Equations \ref{eq:pr} and \ref{eq:r_approx}. Table~\ref{tb:r_stats} also includes redshift information}. {Table~\ref{tab:raw_quadrants} (in Appendix \ref{ap:science_circles}) shows the properties of the science aperture and the X-ray and radio data for each source. Table~\ref{tb:on_off_b_counts} shows the on-axis, off-axis and background counts.} 

\begin{figure*}[htb!]
    \centering
    \includegraphics[width=\textwidth]{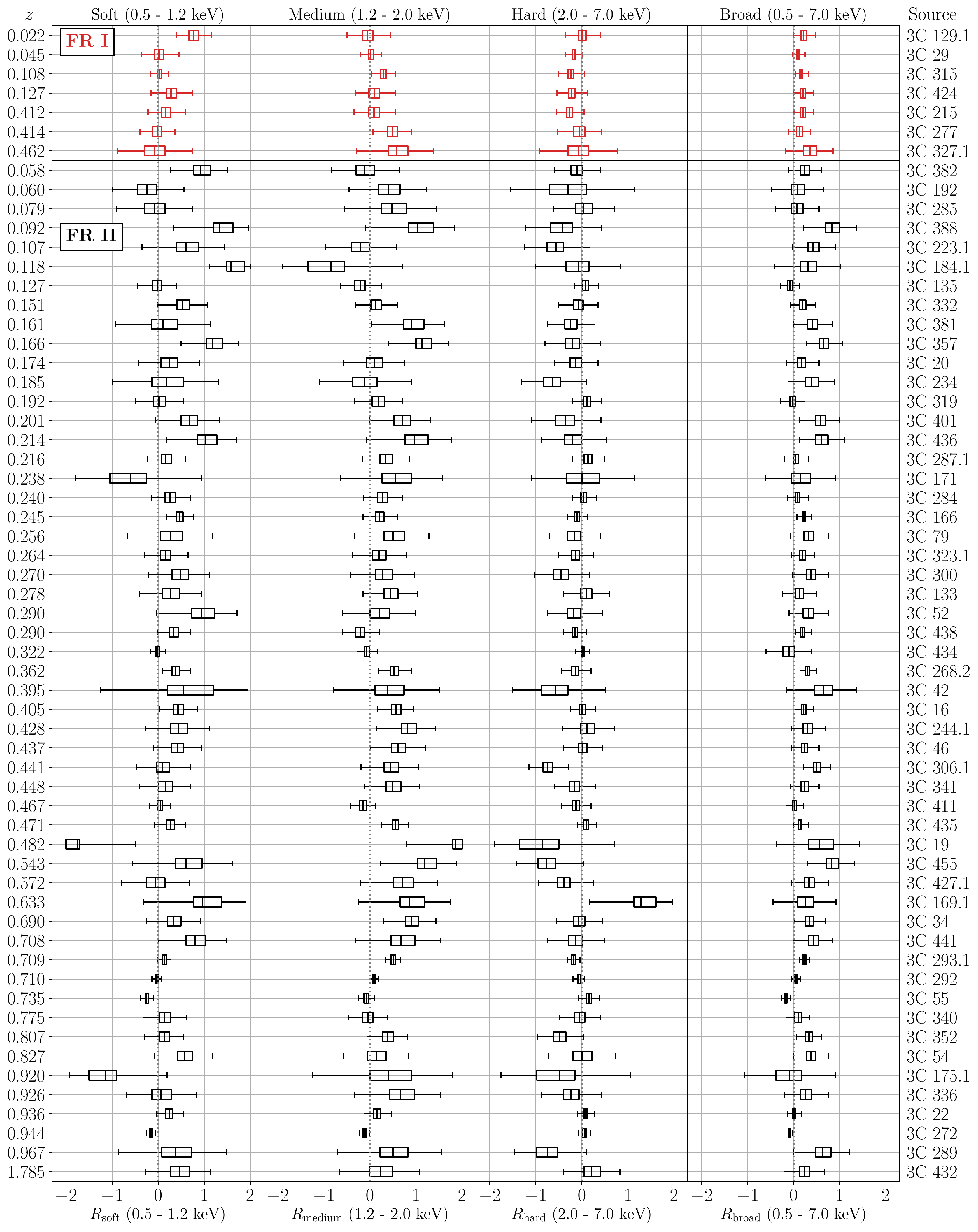}
    \caption{{The box plots of $R$ values for the 60 3CR extragalactic radio sources in our sample. See Figure~\ref{fig:r_3c22_soft} for the box plot definitions.}}
    \label{fig:r_table}
\end{figure*}

{Figure~\ref{fig:r_table} shows the box plots of the median $R$ values for the different bands with the results sorted by FR type and redshift for the sources in our sample, using the aforementioned information. This figure provides an overall birds-eye view of the $R$ distribution functions, their skewness, and behavior with $z$. As noted in \textsection{}\ref{sec:asymmetric}, the distributions become more Gaussian-like (less skewed) with increasing counts, as is evident when comparing with Table~\ref{tb:on_off_b_counts}. This is particularly evident in the broad band, which has the highest counts.}

\begin{figure*}[htb!]
    \centering
    \includegraphics[width=45em]{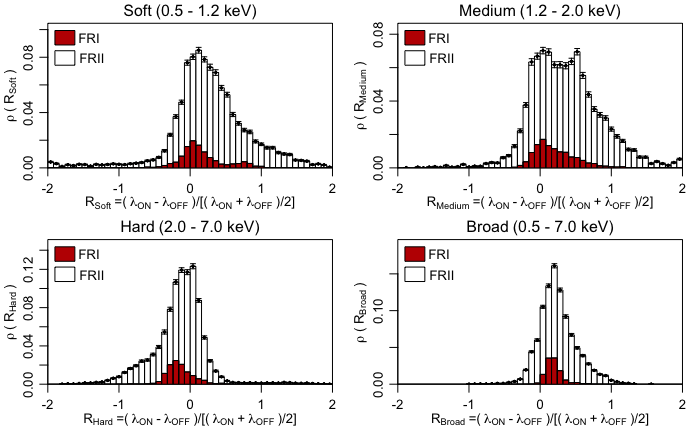}
    \caption{FR I and FR II stacked and normalized histograms of binned $R$ distributions for 
    {7} FR I and 
    {53} FR II sources 
    {, where $\rho(R)$ is the probability density}.
    {The histograms are generated by taking a random sample from each of the $R$ distribution functions, in a given band, and binning the results in an iterative fashion, until the error in frequency is minimized. The tiny error-bars on top of each histogram-bar indicate the bin uncertainties after 250 iterations. Examples of individual $R$ distribution functions are given in Figures~\ref{fig:r_3c401_broad} and~\ref{fig:r_3c22_soft}.}
    }
    \label{fig:r_hists}
\end{figure*}

\begin{figure*}[htb!]
    \centering
    \includegraphics[width=45em]{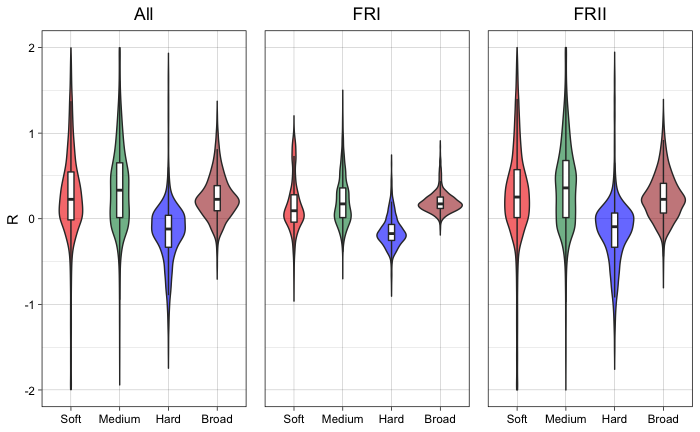}
    \caption{{Summary violin plots of Figure~\ref{fig:r_hists}, where the embedded box-plot median-$R$ values with their $1^{\rm st}$ (Q1) and $3^{\rm rd}$ (Q3) quartile ranges are given in Table~\ref{tb:r_probs}. The lower and upper whiskers are given by the usual $\rm{Q1-(1.5 \times IQR)}$ and $\rm{Q3+(1.5\times IQR)}$, respectively, where IQR is the Interquartile Range ({i.e.}, $\rm{Q3 - Q1}$). For both FR I and FR II sources, in general, it can be seen that the soft, medium, and broad bands favor $R > 0$, whereas the hard band favors $R < 0$.}}
    \label{fig:r_boxplots}
\end{figure*}

{Figure~\ref{fig:r_hists} shows the histograms of the $R$ distributions for different bands. {This is essentially a projection of the distribution functions of Figure~\ref{fig:r_table}, collapsed along $z$: {i.e.}, no correlation is assumed. Here, the skewness represents a general broadening of the $R$-distributions about zero, with small $|R|>0$ outlier-peaks. The broad band gives the overall average trend in $R$,  from which we can ascertain, a slight ON-bias in the soft,  medium bands, and broad bands, and a possible slight OFF-bias in the hard band. Figure~\ref{fig:r_boxplots}} provides summary violin plots of the $R$ values from Figure~\ref{fig:r_hists}. The latter two figures emphasize the overall trend of the $R$ distribution functions. Again,  there is a general trend favoring $R > 0$ for the soft, medium, and broad bands, and $R < 0$ for the hard band. Table~\ref{tb:r_probs} provides a summary of median $R$ values, within the $1^{\rm st}$ and $3^{\rm rd}$ quartile bounds, quantifying this trend.} 

{Table \ref{tab:r_fractions} shows the fraction of sources in the sample for which $R>0$ or $R<0$, within 68.27\% HPD ($\pm1\hat\sigma$), extracted from Table~\ref{tb:r_stats}. Table \ref{tab:r_fractions} represents a fraction of the sample for which $R$ is statistically different from symmetric emission in the on-jet and the off-jet regions.} 

{Figure \ref{fig:r_z} shows $R$ as a function of redshift for all the sources in our sample. Figure \ref{fig:r_z_avg} shows $\overline{R}$ as a function of redshift in two redshift bins: (i) $0 < z < 0.5$ and (ii) $0.5 < z < 1.0$. $\overline{R}$ is the average of the $R$ values for all sources in that particular redshift bin, and the error bars on $\overline{R}$ are the $\pm1\hat{\sigma}$ error bars of all the sources in that particular redshift bin added in quadrature.} 

\begin{figure*}[htb!]
    \centering
    \includegraphics[width=45em]{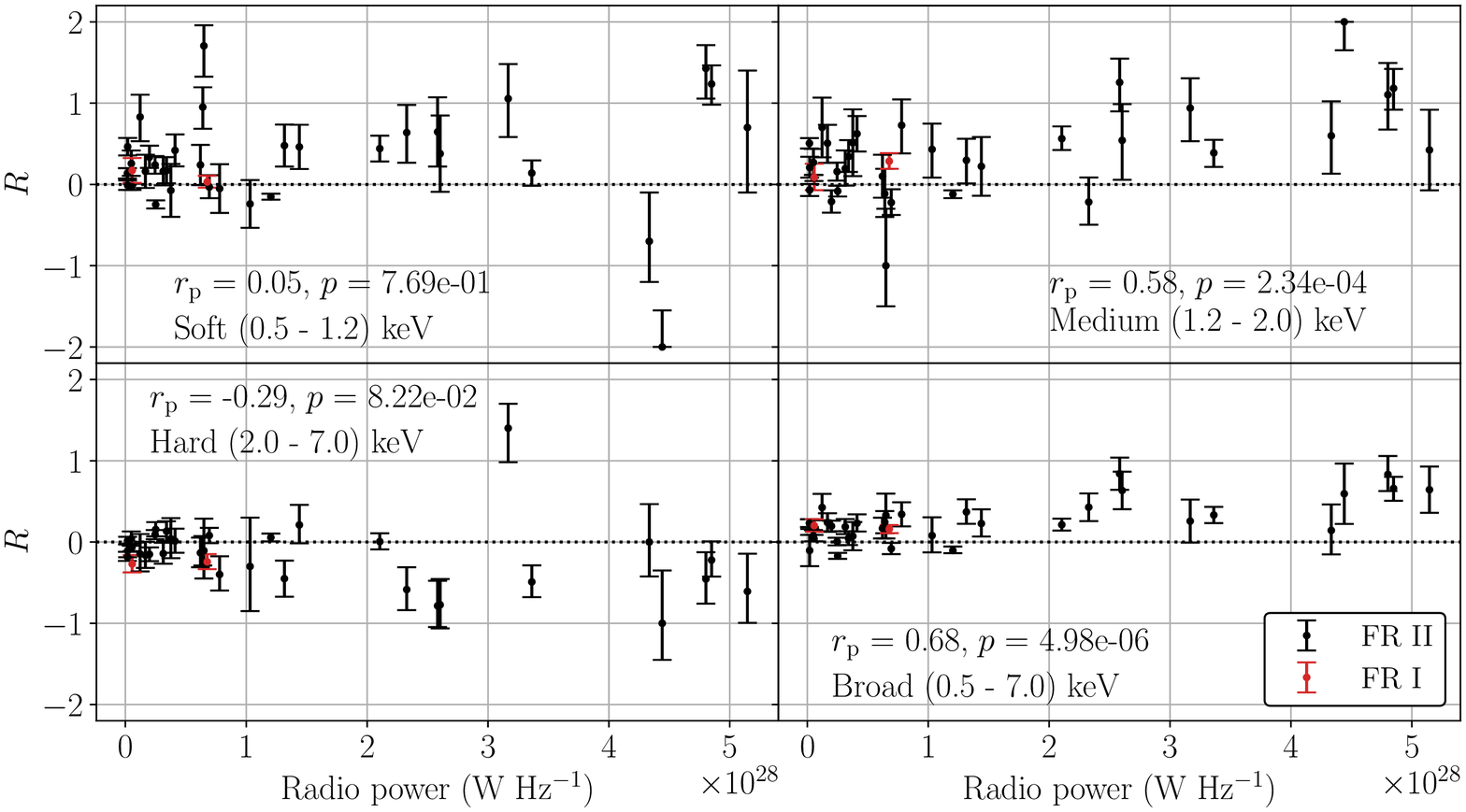}
    \caption{{$R$ as a function of radio power.}}
    \label{fig:r_power}
\end{figure*}

{Figure \ref{fig:r_power} shows $R$ as a function of radio power, respectively. The radio powers were calculated for the sources in our sample that have known 178 MHz fluxes taken from the 3CRR Extragalactic Database.\footnote{\texttt{https://3crr.extragalactic.info/cgi/database}} The radio power was calculated as}

\begin{equation}
    \mbox{$L_{\rm radio} = F_{178\, \rm MHz} \times 4 \pi D_{\rm L}^{2} \times (1 + z)^{-(1 + \alpha)}$}
\label{eqn:L_radio}
\end{equation}

\noindent {where $L_{\rm radio}$ is in units of W Hz$^{-1}$, $F_{178\,\rm MHz}$ is the flux density at 178 MHz in units of Jansky or erg s$^{-1}$ cm$^{-1}$ Hz$^{-1}$, $D_{\rm L}$ is the luminosity distance, and we used a power law index, $\alpha = -\,0.7$.}

\begin{deluxetable}{llll}
   \centering
   \tablecolumns{4}
   \tablewidth{\columnwidth}
   \tablecaption{The median $R$ values from Figure~\ref{fig:r_boxplots} with minus $25^{\rm th}$ (Q1) and plus $75^{\rm th}$ (Q3) percentile bounds.\label{tb:r_probs}}
   \tablehead{\colhead{Band} & \colhead{$R_{\rm combined}$} & \colhead{$R_{\rm FR I}$} & \colhead{$R_{\rm FR II}$}}
   \startdata
        Soft & $\def\arraystretch{0.6}+0.23\!\begin{array}{l}+0.32\\-0.24\end{array}$ & $\def\arraystretch{0.6}+0.09\!\begin{array}{l}+0.19\\-0.13\end{array}$ & $\def\arraystretch{0.6}+0.25\!\begin{array}{l}+0.32\\-0.24\end{array}$ \\ 
      Medium & $\def\arraystretch{0.6}+0.33\!\begin{array}{l}+0.32\\-0.32\end{array}$ & $\def\arraystretch{0.6}+0.17\!\begin{array}{l}+0.19\\-0.16\end{array}$ & $\def\arraystretch{0.6}+0.36\!\begin{array}{l}+0.32\\-0.35\end{array}$ \\ 
        Hard & $\def\arraystretch{0.6}-0.12\!\begin{array}{l}+0.16\\-0.21\end{array}$ & $\def\arraystretch{0.6}-0.17\!\begin{array}{l}+0.11\\-0.08\end{array}$ & $\def\arraystretch{0.6}-0.09\!\begin{array}{l}+0.16\\-0.24\end{array}$ \\ 
       Broad & $\def\arraystretch{0.6}+0.23\!\begin{array}{l}+0.16\\-0.13\end{array}$ & $\def\arraystretch{0.6}+0.17\!\begin{array}{l}+0.08\\-0.05\end{array}$ & $\def\arraystretch{0.6}+0.23\!\begin{array}{l}+0.19\\-0.16\end{array}$ \\ 
   \enddata
\end{deluxetable}

\begin{deluxetable*}{l|ll|ll|ll}
\centering
\tablecolumns{7} 
\tablewidth{700pt}
\tabletypesize{\scriptsize}
\tablecaption{{Fraction of sources in our sample for which $R > 0$ or $R <0$, within 68.27\% HPD ($\pm1\hat{\sigma}$).}}
\label{tab:r_fractions}
\tablehead{ 
\colhead{Band} & \colhead{$N_{\rm{both}}(R\pm1\hat{\sigma}>0)$} & \colhead{$N_{\rm{both}}\,(R\pm1\hat{\sigma}<0$)} & \colhead{$N_{\rm{FRI}}\,(R\pm1\hat{\sigma}>0$)} & \colhead{$N_{\rm{FRI}}\,(R\pm1\hat{\sigma}<0$)} & \colhead{$N_{\rm{FRII}}\,(R\pm1\hat{\sigma}>0$)} & \colhead{$N_{\rm{FRII}}\,(R\pm1\hat{\sigma}<0$)}
}
\startdata 
Soft   & 31/60 (51.7\%)             & 5/60 (8.3\%)            & 3/7 (42.9\%)               & 0/7 (0.0\%)             & 28/53 (52.8\%)             & 5/53 (9.4\%)            \\
Medium & 36/60 (60.0\%)             & 6/60 (10.0\%)           & 3/7 (42.9\%)               & 0/7 (0.0\%)            & 33/53 (62.3\%)             & 6/53 (11.3\%)           \\
Hard   & 6/60 (10.0\%)               & 24/60 (40.0\%)          & 0/7 (0.0\%)                & 4/7 (57.1\%)            & 6/53 (11.3\%)              & 20/53 (37.7\%)          \\
Broad  & 46/60 (76.7\%)             & 3/60 (5.0\%)        & 7/7 (100.0\%)               & 0/7 (0.0\%)             & 39/53 (73.6\%)             & 3/53 (5.7\%)    
\enddata 
\end{deluxetable*}

\section{Discussion} \label{sec:discussion}
{The broad band ($0.5 - 7.0$ keV) has the highest detected flux (and therefore the highest signal-to-noise ratio and the lowest error bars in $R$), so we focus on the broad band results for the physical interpretation.
Table \ref{tab:r_fractions} shows that a majority ($\sim 77\%)$ of the sources in our combined FR I and FR II sample have $R \pm 1\hat\sigma > 0$ in the broad band, suggesting that the non-thermal X-ray emission from the radio lobes typically dominates the thermal X-ray emission from the hot gas in the environments. On the other hand, the thermal emission from the ICM dominates the non-thermal emission from the lobes for only $5\%$ of the sources in our combined sample.}  

{For the FR I (FR II) subsample, the on-jet axis non-thermal emission dominates the off-jet axis thermal emission for 100$\%$ ($\sim 74\%)$ of the sources, whereas the off-axis thermal emission dominates for 0\% ($\sim 6\%$) of the sources.} 


{The flux due to thermal bremsstrahlung emission from the hot gas in the off-jet axis regions in the ICM environments of radio galaxies can be expected to decrease with increasing redshift. On the other hand, the flux due to the IC scattering of the CMB photons by relativistic electrons in the radio lobes in the on-jet axis regions is roughly redshift independent due to the increasing CMB energy density with increasing redshift.} 

\begin{figure*}[htb!]
    \centering
    \includegraphics[width=45em]{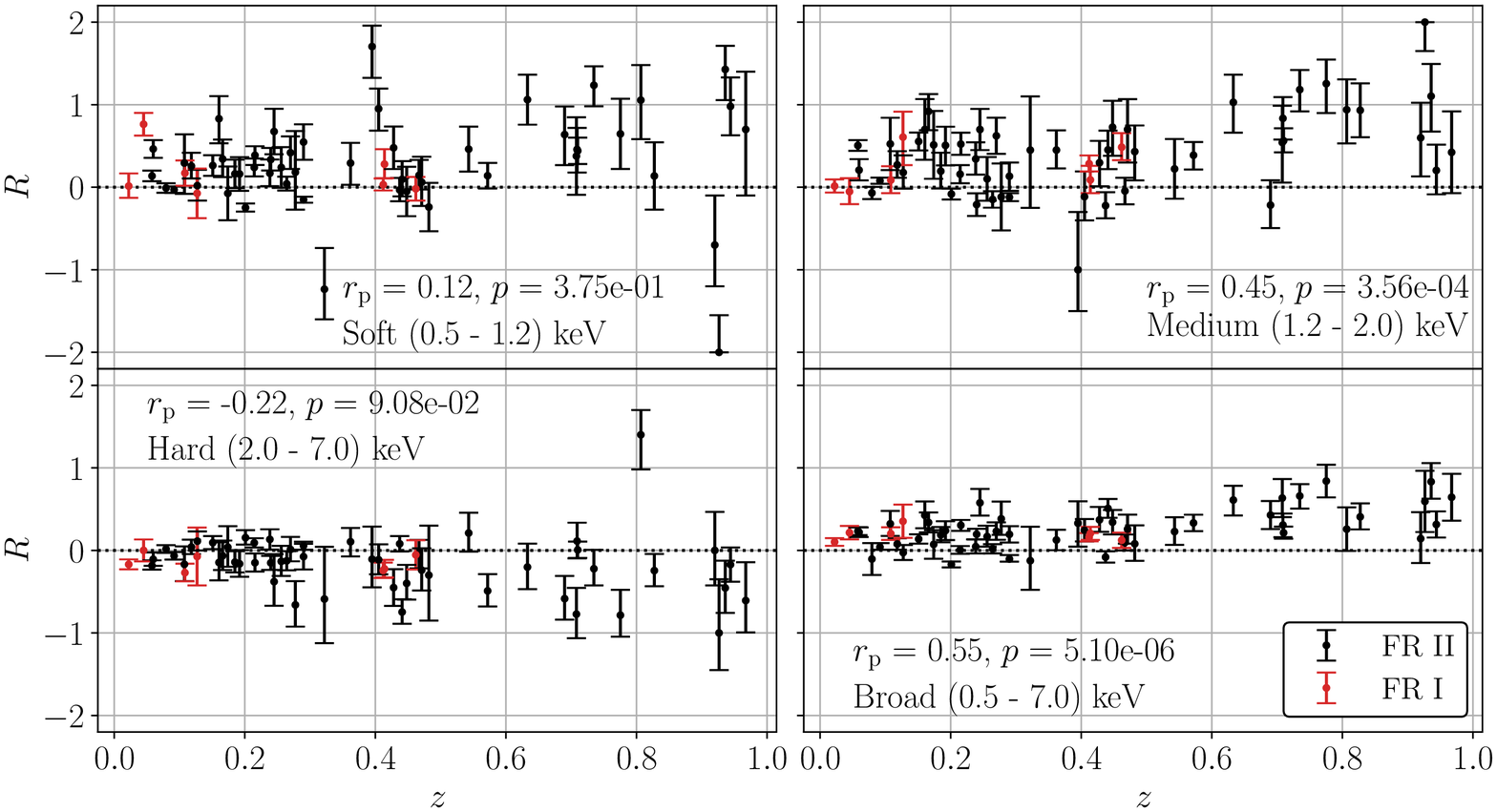}
    \caption{{$R$ as a function of redshift.}}
    \label{fig:r_z}
\end{figure*}

\begin{figure*}[htb!]
    \centering
    \includegraphics[width=45em]{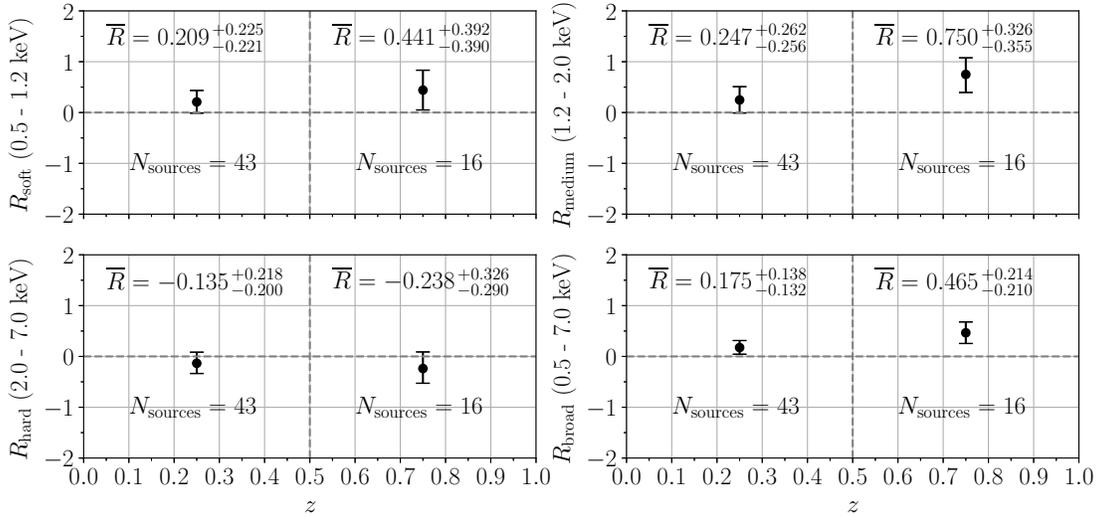}
    \caption{{$\overline{R}$ as a function of redshift in two redshift bins: (i) $0 < z < 0.5$ and (ii) $0.5 < z < 1.0$. The source 3C 432, which has a redshift of 1.785, was excluded in these calculations in order to simplify the redshift bins, since 3C 432 is the only source in our sample with redshift greater than 1. Here, $\overline{R}$ is the average of the $R$ values for all sources in that particular redshift bin, and the errorbars on $\overline{R}$ are the $\pm1\hat{\sigma}$ errorbars of all the sources in that particular redshift bin added in quadrature. The broad band has the highest signal-to-noise ratio and hence the smallest errorbars. $\overline{R}$ tends to increase with $z$ in the broad band, suggesting that the on-axis jet emission is due to the inverse Compton scattering of CMB photons by hot electrons, a process for which the observed flux is roughly redshift independent due to the increasing CMB energy density with $z$.}}
    \label{fig:r_z_avg}
\end{figure*}

{Figures \ref{fig:r_z} and \ref{fig:r_z_avg} show that $R$ tends to increase with redshift in the broad band. This marginal increase is also seen between $R$ and the radio power, which can be expected since (i) radio power depends on redshift (see equation \ref{eqn:L_radio}), and (ii) the selection function of the 3CR catalog, for which the 9 Jy flux density limit  at $z = 1$ corresponds to a power of $\sim 4 \times 10^{28}$ W Hz$^{-1}$. The marginal increase in $R$ with redshift is consistent with the IC/CMB emission mechanism in the on-jet axis regions. Therefore, the fact that (i) $R \pm 1\hat\sigma > 0$ for a majority of the sources in the broad band, and (ii) $R$ tends to increase with redshift suggests that (i) the dominant X-ray emission mechanism in the radio lobes is IC/CMB in origin, and that (ii) the non-thermal emission due to IC/CMB from the radio lobes in general tends to dominate  the thermal bremsstrahlung emission from the hot gas in the ICM environment of  radio galaxies.} 

{There have been various studies suggesting that the X-ray emission mechanism in jets is likely due to the IC upscattering of CMB photons. \citet{Tavechhio2000} and \citet{Celotti} used \textit{Chandra} observations of the 100 kpc jet of PKS 0637-752 to propose the IC/CMB mechanism for the jet X-ray emission. \citet{harris2002} presented a formulation of anisotropic IC emission from relativistically moving jets, suggesting that the jets have high bulk relativistic velocities on kpc scales, and that higher redshift sources require lower IC beaming compared to lower redshift sources due to the higher CMB energy density. \citet{croston2005} used archival \textit{Chandra} and \textit{XMM-Newton} data for a sample of 40 FR II radio galaxies and quasars between $0.05 < z < 2$, and found that the X-ray emission from the radio lobes can be attributed to the IC/CMB mechanism. \citet{Kharb12a} used deep \textit{Chandra} and \textit{HST} observations of two quasars, PKS B0106+013 and 1641+399 (3C 345), and found that the X-rays from the jets are produced from IC mechanism from relativistically boosted CMB seed photons.  \citet{ineson_2017}  measured the X-ray lobe IC emission for a sample of 37 FR II galaxies between $0.1 < z < 0.5$ to investigate the internal radio lobe conditions and lobe dynamics.  \citet{Mingo} studied the X-ray emission associated with the extended structures (jets, lobes, and environments) of 26 sources from the 2 Jy sample of radio galaxies\footnote{https://2jy.extragalactic.info/2Jy\_home\_page.html} between $0.05 < z < 0.2$, although the FR II lobe analysis in \citet{Mingo} was part of the wider FR II lobe study of \citet{ineson_2017}.} 

{More recently, \citet{jimenezgallardo2020extended} studied the extended emission in the $0.5 - 3.0$ keV band with \textit{Chandra} data around 35 FR II radio galaxies between $0.05 < z < 0.9$ selected from the 3CR catalog. About half of their sample overlaps with ours. They found that the non-thermal IC/CMB process dominates in $\sim 70\%$ of the sources in their sample, while the thermal emission from the ICM dominates in $\sim 15\%$ of them. As expected, the results in our study are in general agreement with those of  \citet{jimenezgallardo2020extended}. }

{Thus, our results are consistent with previous work which suggests that the X-ray emission from radio lobes is due to IC scattering of the CMB. }





\section{Sources with X-ray emission associated with radio hotspots} \label{sec:report}

During this study, we found some sources with X-ray emission spatially correlated with the radio hotspots. Since in our selection criteria we did not include sources with X-ray emission associated with the radio hotspots, the sources listed below were excluded from the sample.

\begin{tabular}{ll}
3C 41: & Southeast hotspot.\\
3C 153: & Northeast hotspot.\\
3C 263.1: & Northeast and southwest hotspots.\\
3C 337: & Southeast hotspot.
\end{tabular}

\noindent We also confirm the presence of X-ray emission associated with the radio hotspots of the following sources previously reported in \cite{arXiv:1609.07145}.

\begin{tabular}{ll}
3C 299: & Northeast and southwest hotspots.\\
3C 325: & Northwest and southeast hotspots.\\
3C 349: & Southeast hotspot.
\end{tabular}

\begin{figure}[htb!]
	\centering
	\includegraphics[width=\columnwidth]{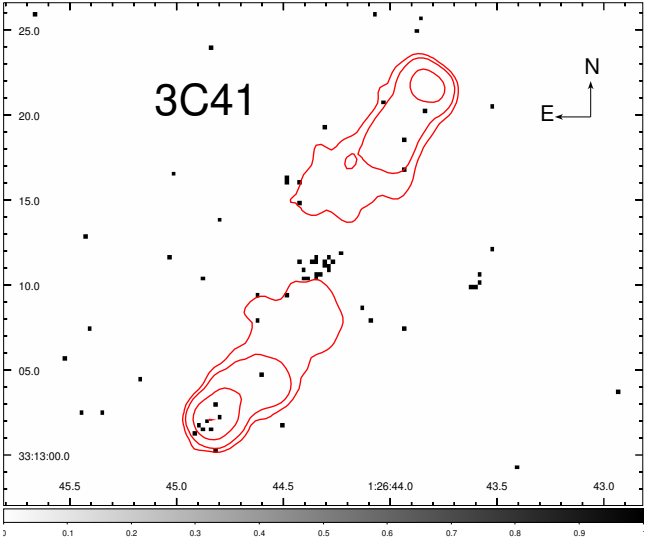}
	\caption{3C 41 broad band X-ray image with radio contours in red. X-ray and radio emission spatial correlation in south-east hotspot.}
	\label{fig:3C41}
\end{figure}

\begin{figure}[htb!]
	\centering
	\includegraphics[width=\columnwidth]{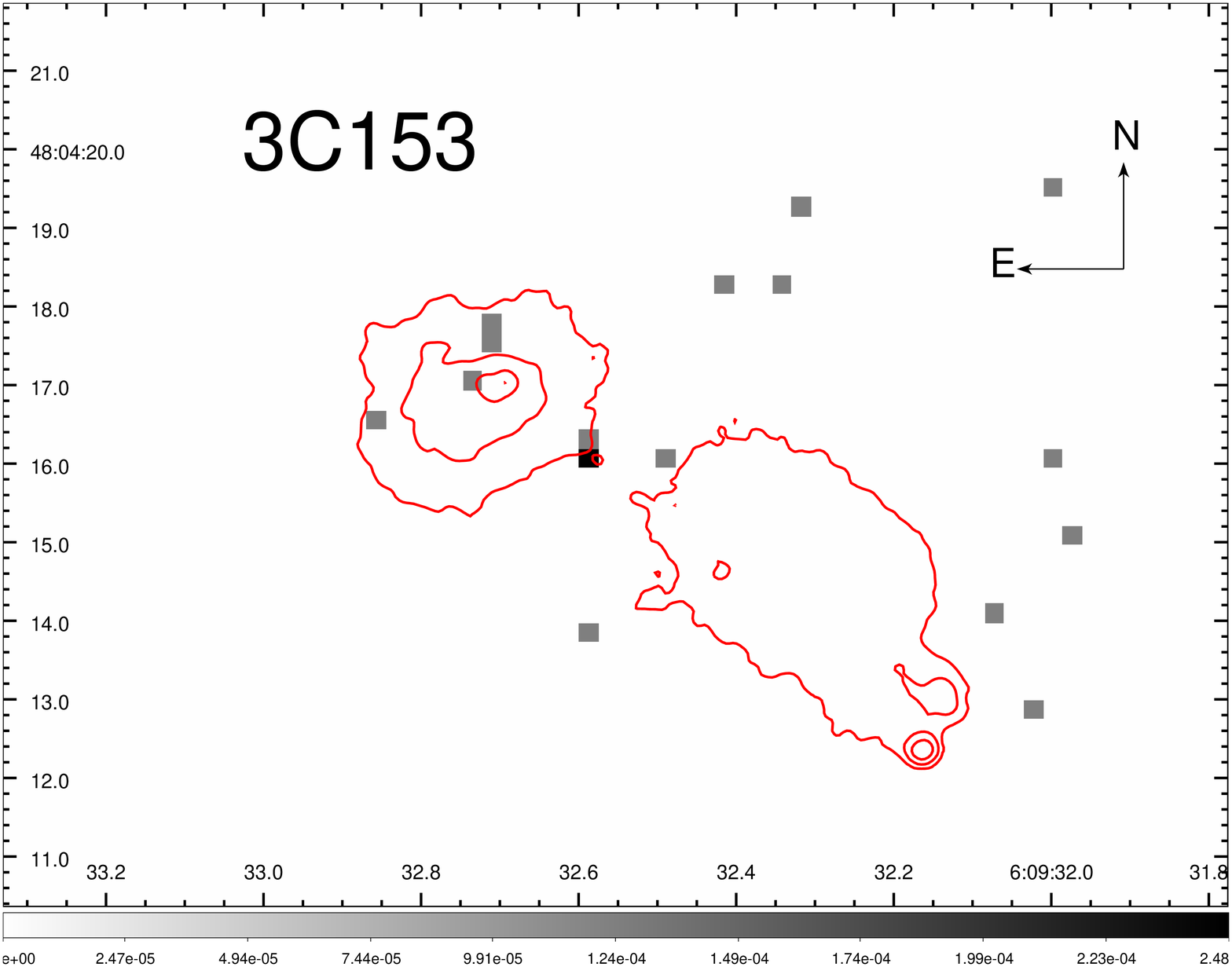}
	\caption{3C 153 broad band X-ray image with radio contours in red. X-ray and radio emission spatial correlation in north-east radio hotspot.}
	\label{fig:3C153}
\end{figure}

\begin{figure}[htb!]
	\centering
	\includegraphics[width=\columnwidth]{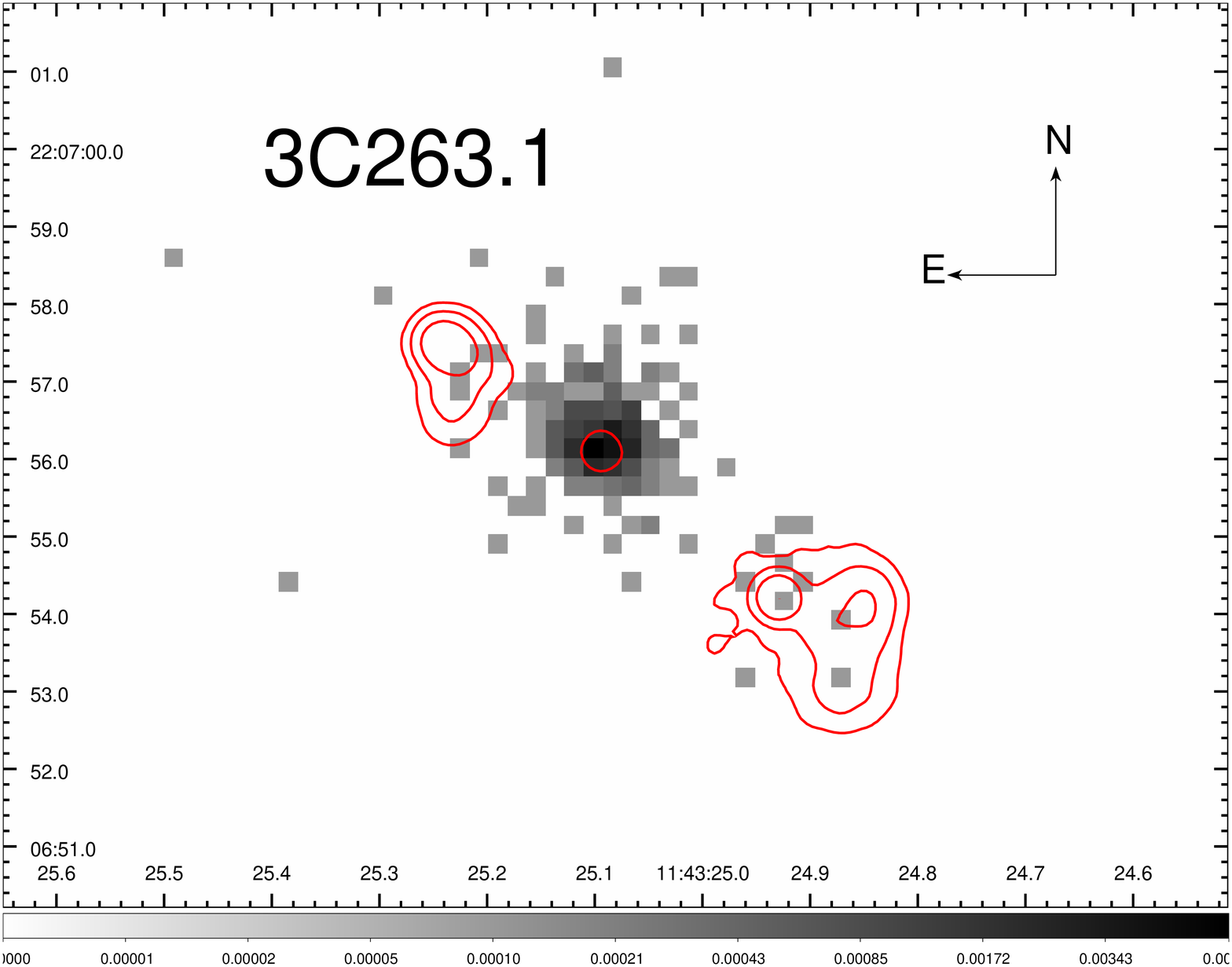}
	\caption{3C 263.1 broad band X-ray image with radio contours in red. X-ray and radio emission spatial correlation in north-east and south-west radio hotspots.}
	\label{fig:3C263.1}
\end{figure}

\begin{figure}[htb!]
	\centering
	\includegraphics[width=\columnwidth]{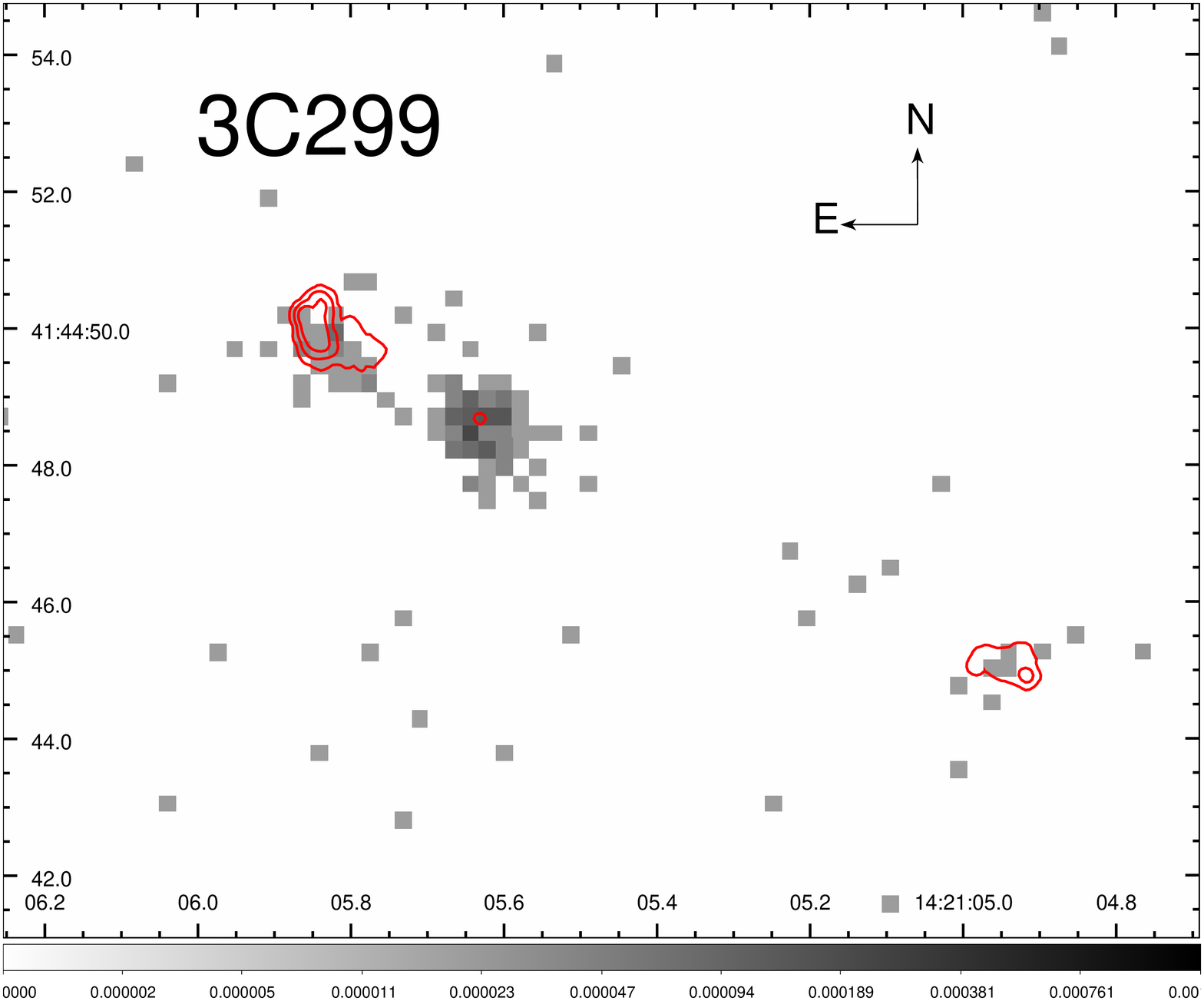}
	\caption{3C 299 broad band X-ray image with radio contours in red. X-ray and radio emission spatial correlation in north-east and south-west radio hotspots.}
	\label{fig:3C299}
\end{figure}

\begin{figure}[htb!]
	\centering
	\includegraphics[width=\columnwidth]{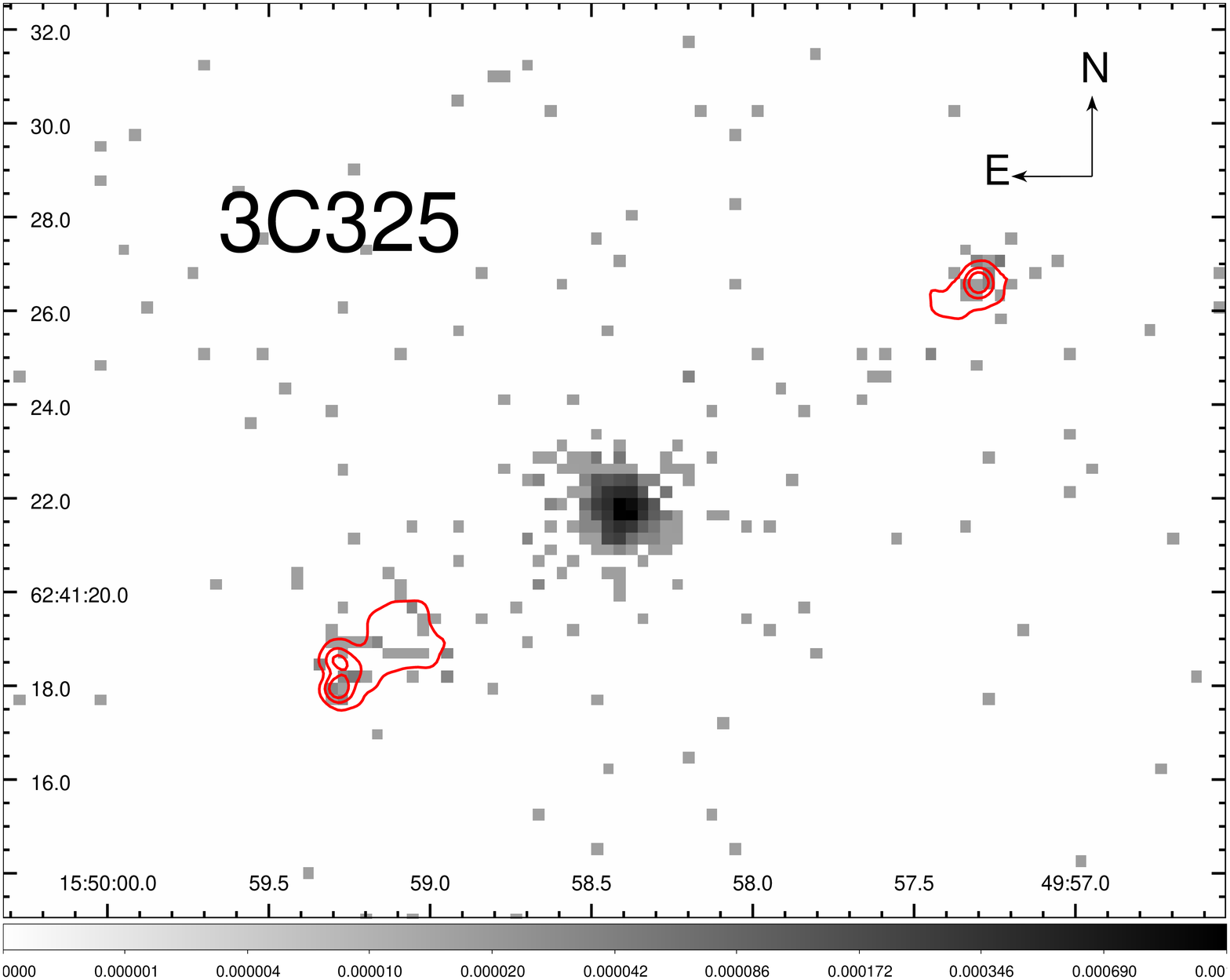}
	\caption{3C 325 broad band X-ray image with radio contours in red. X-ray and radio emission spatial correlation in north-west and south-east radio hotspots.}
	\label{fig:3C325}
\end{figure}

\begin{figure}[htb!]
	\centering
	\includegraphics[width=\columnwidth]{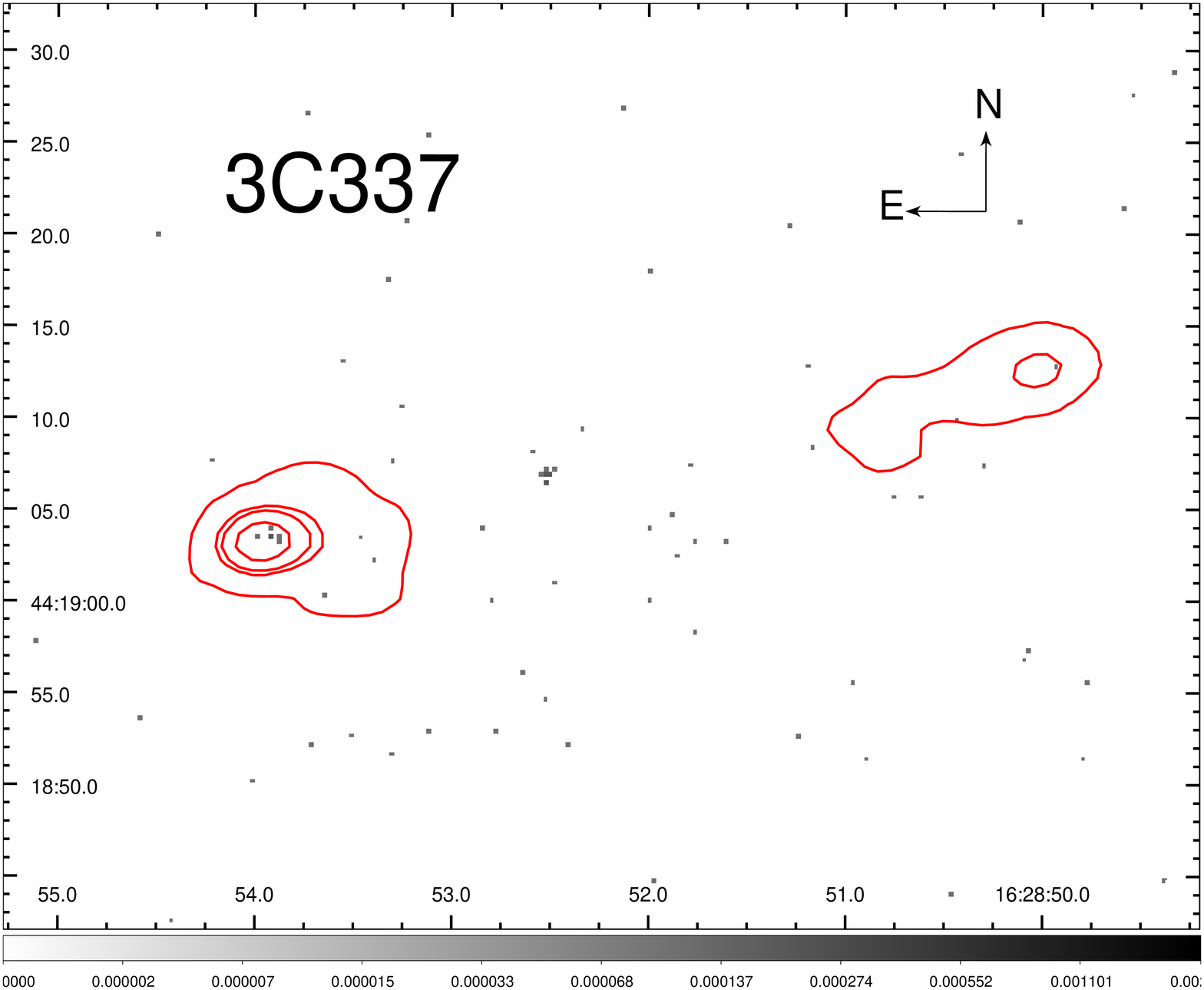}
	\caption{3C 337 broad band X-ray image with radio contours in red. X-ray and radio emission spatial correlation in south-east radio hotspot.}
	\label{fig:3C337}
\end{figure}

\begin{figure}[htb!]
	\centering
	\includegraphics[width=\columnwidth]{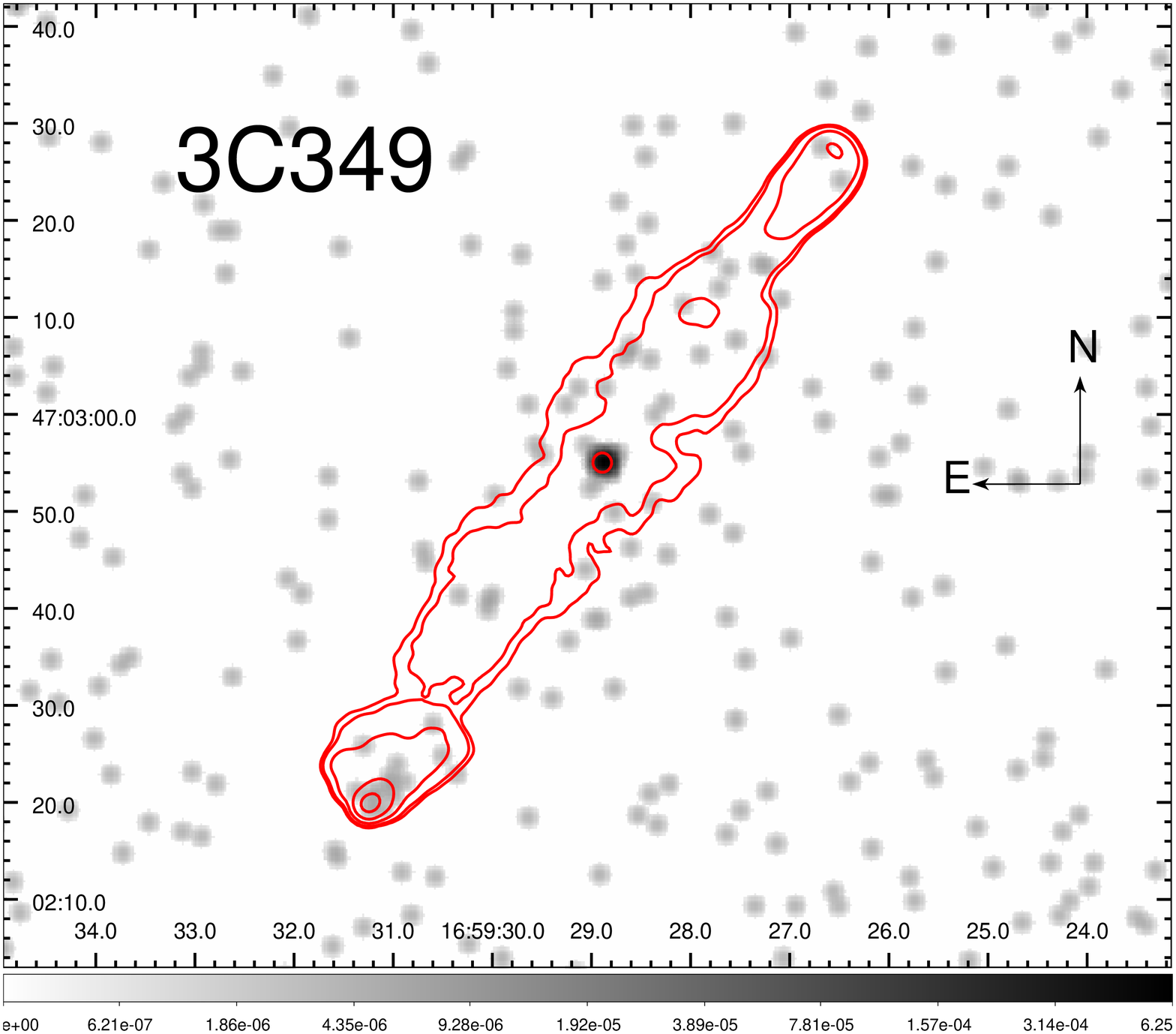}
	\caption{3C 349 broad band X-ray image smoothed with a Gaussian. Radio contours in red. X-ray and radio emission spatial correlation in south-east radio hotspot.}
	\label{fig:3C349}
\end{figure}

\noindent These sources were also excluded from our sample. Figures \ref{fig:3C41}-\ref{fig:3C349} show the broadband X-ray images with overlaid radio contours of the sources mentioned above. 

\section{Summary}

{This paper studied the extended faint X-ray emission associated with the radio lobes and the hot gas in the ICM environment for a sample of 60 radio galaxies (7 FR I and 53 FR II) selected from the 298 extragalactic radio sources identified in the 3CR catalog. We looked for any asymmetry in the extended faint X-ray emission between regions that contain the radio lobes and regions in the hot environment that do not contain the radio lobes. To do so, we constructed a circular science aperture for each source with four quadrants of equal area, where two quadrants contain the radio lobes (on-jet axis) and the two other quadrants do not contain the radio lobes and are in the environment regions (off-jet axis). We used shallow ($\sim 10$ ks) archival X-ray observations from \textit{Chandra} for our sample in four energy bands: soft ($0.5 - 1.2$ keV), medium ($1.2 - 2.0$ keV), hard ($2.0 - 7.0$ keV), and broad ($0.5 - 7.0$ keV). The radio images were taken with the VLA.}

{We used Bayesian techniques on the Poission X-ray counts in the on and off-jet axis regions to determine whether the on-jet axis or the off-jet axis emission dominates for each source and each band. To do so, we used the parameter $R$, defined as $R = (S_{\rm ON} - S_{\rm OFF}) \,/\,\, [(S_{\rm ON} + S_{\rm OFF})\,/\,2]$, such that $R > 0$ corresponds to higher on-jet axis emission (from the radio lobes) and $R < 0$ corresponds to higher off-jet axis emission (from the hot gas in the environment). Since the broad band had the most detected photons and therefore the highest signal-to-noise ratio, we focus our physical interpretation using the broad band results.} 

{We found that the non-thermal X-ray emission from the radio lobes dominates the thermal X-ray emission from the hot gas in the ICM for a majority ($\sim 77\%$) of sources in the combined FR I and FR II sample. The thermal X-ray emission from the ICM dominates the non-thermal X-ray lobe emission for only $\sim 5\%$ of the sources in the combined sample.}

{We also found that $R$ in the broad band has a marginal increase with increasing redshift. The observed flux due to the off-jet axis thermal bremsstrahlung emission from the hot gas in the environment can be expected to decrease with increasing redshift. On the other hand, the observed flux due to the on-jet axis inverse Compton scattering of CMB seed photons by relativistic electrons in the radio lobes is roughly independent of redshift, due to the increasing CMB energy density at increasing redshift. The fact that $R$ tends to increase with increasing redshift suggests that dominant X-ray emission mechanism in the radio lobes is IC/CMB in origin.   Therefore, our results suggest that the non-thermal X-ray emission due to the IC/CMB process from the radio lobes in general tends to dominate the thermal X-ray emission from the hot gas in the ICM environment of radio galaxies. Our results are  consistent with previous work. }

\section{Acknowledgments}
{We thank the referee for detailed comments which helped to improve the paper}.
We thank Dr. Francesco Massaro for his comments and for providing radio images for some sources, and Dr. Adrian Vantyghem for his feedback regarding interpretation of the Bayesian analysis results.  This work used archival data from $Chandra$ and software provided by the $Chandra$ X-ray Center (CXC) in the application package CIAO. \texttt{SAOImage DS9} was used for the analysis of the X-ray and radio data. \texttt{SAOImage DS9} was developed by the Smithsonian Astrophysical Observatory \citep{ds9}. Many radio images were downloaded from NVAS. The National Radio
Astronomy Observatory is operated by Associated Universities, Inc., under contract with the National Science Foundation. The research for this paper involved the use of NASA's Astrophysical Data System and the NASA/IPAC Extragalactic Database (NED) provided by the Jet Propulsion Laboratory, California Institute of Technology, under contract with the National Aeronautics and Space Administration. The research of Drs. S. Baum, M. Boyce,  and C. O'Dea was supported by the Natural Sciences and Engineering Research Council (NSERC)  of Canada.

\clearpage
\appendix
 \section{Bayesian Analysis}
\label{ap:bayesian}

{Application of Bayesian analysis in astronomy is rapidly expanding in its areas of influence, and, as such, is useful to have in one's toolkit. However, it can have a steep learning curve. In the area of X-ray astronomy, dealing with low statistics, authors, such as, \cite{dyk2001} and \cite{park2006} have paved the way, providing theory and application examples. In this appendix, their work is extended by providing a case study of the prior distribution function, Equation (\ref{eq:pr}), for the jet-axes ratio, $R$, Equation (\ref{eq:r}).}

In the derivation of Equation (\ref{eq:pr}), the source, $S_x$, and background, $B_x$, counts are modeled in terms of their expected source, $\lambda_x$, and background, $\xi_x$, counts, respectively, as

\begin{equation}
    \left.\begin{array}{l@{\;\stackrel{d}{\sim}\;}l}
         S_x & \;\text{Poisson}(e_C\,(\lambda_x+\xi_x)) \\
         B_x & \;\text{Poisson}(r\,e_C\,\xi_x)
    \end{array}\right\}\,,\label{eq:fish}
\end{equation}

\noindent where $x$ = ON, OFF, $r=2$ (see discussion in Section \ref{sec:asymmetric}) is a scale factor that accounts for the differences in source and background measurement areas and sensitivities, and $e_C=1$ (see discussion in Section 2.1 of \citet{park2006}) is an instrumental parameter that accounts for the variations in the soft ($e_S$), medium ($e_M$), hard ($e_H$), and broad ($e_B$) band (i.e., $C=S,M,H,B$) effective areas, exposure times, etc. This approach is justified, as the sum of two independent Poisson random variables (\textit{i.e.}, $\lambda_x$ and $\xi_x$) is a Poisson random variable  (\textit{i.e.}, $\lambda_x+\xi_x$) \citep{hoff2010}, and so we may express Equation (\ref{eq:r}) as

\begin{equation}
    R = \frac{\lambda_{\rm ON}-\lambda_{\rm OFF}}{(\lambda_{\rm ON}+\lambda_{\rm OFF})/2}\,.
    \label{eq:expected}
\end{equation}

\begin{figure}[htb!]
\centering
\begin{tabular}{l@{$\;\;\;\;\;\;\;\;$}r}
    \includegraphics[width=0.47\linewidth]{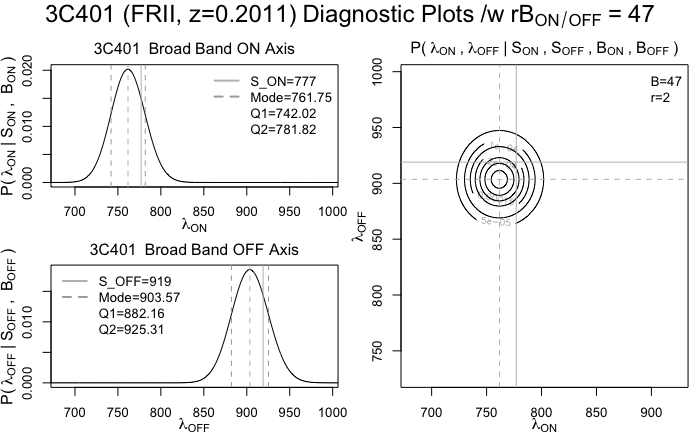} & 
    \includegraphics[width=0.47\linewidth]{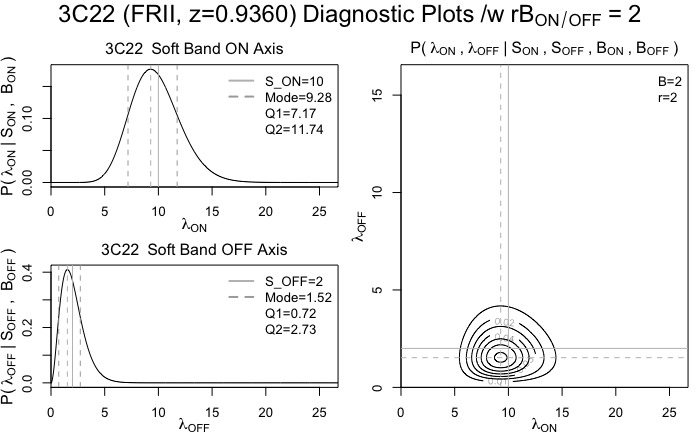}
\end{tabular}
\caption{Example plots of Equation (\ref{eq:marginal}) for sources 3C 401 (left panel) and 3C 22 (right panel) for the broad and soft bands, respectively, with input values from Table~\ref{tb:on_off_b_counts}. The probability distributions $P(\lambda_{\rm ON}|S_{\rm ON},B_{\rm ON})$, $P(\lambda_{\rm OFF}|S_{\rm OFF},B_{\rm OFF})$, and $P(\lambda_{\rm ON},\lambda_{\rm OFF}|S_{\rm ON},S_{\rm OFF},B_{\rm ON},B_{\rm OFF})$ are plotted in the top-left, bottom-left, and right sub-plots of the panels, respectively. {The grey solid-lines show input $S_{\rm ON/OFF}$ values, and the dashed-lines show the expected $\lambda_{\rm ON/OFF}$ mode-values with 68.27\% HPD ($\sim1\sigma$) boundaries.}}
\label{fig:diagnostic_plots}
\end{figure}

The joint posterior distribution of $\lambda_x$ and $\xi_x$, given prior knowledge $S_x$ and $B_x$, can be expressed using the Bayes' Theorem \citep{park2006} as
\begin{equation}
    P(\lambda_x,\,\xi_x|S_x,B_x)=\frac{P(\lambda_x,\,\xi_x)\,P(S_x,B_x|\lambda_x,\,\xi_x)}{\int\int P(\lambda^\prime_x,\,\xi^\prime_x)\,P(S_x,B_x|\lambda^\prime_x,\,\xi^\prime_x)\,d\lambda^\prime_x\,d\xi^\prime_x}\,.
    \label{eq:bayes}
\end{equation}

\noindent where $P(S_x,B_x|\lambda_x,\,\xi_x)=P(S_x|\lambda_x,\,\xi_x)\,P(B_x|\xi_x)$ (see Equation (\ref{eq:fish})) is the likelihood of observing $S_x$ and $B_x$ given $\lambda_x$ and $\xi_x$, and $P(\lambda_x,\,\xi_x)$ is our prior probability distribution function. In order for the posterior distribution to have the same parametric form as the likelihood functions, Equation (\ref{eq:fish}) $\stackrel{d}{\sim}$ Poissonian, we model $\lambda_x$ and $\xi_x$ using conjugate gamma-prior distributions as 

\begin{equation}
    \left.\begin{array}{l@{\;\stackrel{d}{\sim}\;}l}
         \lambda_x & \;\gamma(\psi_{S_{x_1}},\psi_{S_{x_2}}) \\
         \xi_x & \;\gamma(\psi_{S_{x_3}},\psi_{S_{x_4}})
    \end{array}\right\}\,,
\end{equation}

\noindent with the hyperparameters
\begin{equation}
    \left.\begin{array}{l@{\;=\;}l}
         \psi_{S_{x_1}} & S_x+1 \\
         \psi_{S_{x_2}} & 1 \\
         \psi_{S_{x_3}} & B_x+1\\
         \psi_{S_{x_4}} & 1
    \end{array}\right\}\,,\label{eq:psis}
\end{equation}

\noindent where $\lambda_x$ and $\xi_x$ are assumed to be a priori independent, $i.e.$, $P(\lambda_x,\,\xi_x)=P(\lambda_x)\,P(\xi_x)$, \citet{dyk2001}.\footnote{Assuming an informative model: $i.e.$, we have measured background counts, $B_x$.} We note at this point, that the analysis being done here, is similar to that of \citet{park2006}, except they are computing harness ratios

\begin{equation}
    \mathcal{HR}=\frac{\lambda_{\rm H} - \lambda_{\rm S}}{\lambda_{\rm H} + \lambda_{\rm S}}\,,
\end{equation}

\noindent where $\lambda_{\rm H}$ and $\lambda_{\rm S}$ are the expected hard and soft band counts, respectively, which is analogous in form to Equation (\ref{eq:expected}), with the equivalence $\lambda_{\rm ON}\Leftrightarrow\lambda_{\rm H}$ and $\lambda_{\rm OFF}\Leftrightarrow\lambda_{\rm S}$. So comparing Equation (\ref{eq:bayes}) with Equation (A6) of \citet{park2006}, we obtain the analytical expression

\begin{equation}
    P(\lambda_x|S_x\,B_x)=\frac{\displaystyle\sum_{\ell=0}^{S_x}\frac{1}{\Gamma(\ell+1)\Gamma(S_x-\ell+1)}\,\frac{\Gamma(S_x-\ell+B_x+\psi_{S_{x_3}})}{(e_C+r\,e_C+\psi_{S_{x_4}})^{S_x-\ell+B_x+\psi_{S_{x_3}}}}\,\lambda_x^{\ell+\psi_{S_{x_1}}-1}\,e^{-(e_C+\psi_{S_{x_2}})\lambda_x}}{\displaystyle \sum_{k=0}^{S_x}\frac{1}{\Gamma(k+1)\Gamma(S_x-k+1)}\,\frac{\Gamma(S_x-k+B_x+\psi_{S_{x_3}})}{(e_C+r\,e_C+\psi_{S_{x_4}})^{S_x-k+B_x+\psi_{S_{x_3}}}}\,\frac{\Gamma(k+\psi_{S_{x_1}})}{(e_C+\psi_{S_{x_2}})^{k+\psi_{S_{x_1}}}}}\,,
\end{equation}

\noindent where the nuisance parameter, $\xi_x$, has been integrated out. Our joint distribution for our expected ON/OFF counts is given by

\begin{equation}
    P(\lambda_{\rm ON},\lambda_{\rm OFF}|S_{\rm ON},S_{\rm OFF},B_{\rm ON},B_{\rm OFF})=P(\lambda_{\rm ON}|S_{\rm ON},B_{\rm ON})\,P(\lambda_{\rm OFF}|S_{\rm OFF},B_{\rm OFF})\,,\label{eq:marginal}
\end{equation}

\noindent assuming $\lambda_{\rm ON}$ and $\lambda_{\rm OFF}$ are a priori independent. Figure (\ref{fig:diagnostic_plots}) shows example plots for sources 3C 401 and 3C 22 for the broad and soft bands, respectively.

Equation (\ref{eq:marginal}) simplifies to

\begin{equation}
    P(\lambda_{\rm ON},\lambda_{\rm OFF}|S_{\rm ON},S_{\rm OFF},B)=P(\lambda_{\rm ON}|S_{\rm ON},B)\,P(\lambda_{\rm OFF}|S_{\rm OFF},B)\,,
\end{equation}

\noindent as $B_{\rm ON}=B_{\rm OFF}=B/r$, Equation (\ref{eq:bees}), and so

\begin{equation}
    P(\lambda_{\rm ON},\lambda_{\rm OFF}|S_{\rm ON},S_{\rm OFF},B)=\sum_{i=0}^{S_{\rm ON}}\sum_{j=0}^{S_{\rm OFF}}N(i,S_{\rm ON},B)N(j,S_{\rm OFF},B)\lambda_{\rm ON}^{\alpha_{\rm ON}(i)}\lambda_{\rm OFF}^{\alpha_{\rm OFF}(j)}e^{-\beta(\lambda_{\rm ON}+\lambda_{\rm OFF})}\label{eq:joint}
\end{equation}

\noindent where

\begin{equation}
    N(\ell,S_x,B)=\frac{\displaystyle\frac{1}{\Gamma(\ell+1)\Gamma(S_x-\ell+1)}\,\frac{\Gamma(S_x-\ell+B+\psi_{S_{x_3}})}{(e_C+r\,e_C+\psi_{S_{x_4}})^{S_x-\ell+B+\psi_{S_{x_3}}}}}{\displaystyle \sum_{k=0}^{S_x}\frac{1}{\Gamma(k+1)\Gamma(S_x-k+1)}\,\frac{\Gamma(S_x-k+B+\psi_{S_{x_3}})}{(e_C+r\,e_C+\psi_{S_{x_4}})^{S_x-k+B+\psi_{S_{x_3}}}}\,\frac{\Gamma(k+\psi_{S_{x_1}})}{(e_C+\psi_{S_{x_2}})^{k+\psi_{S_{x_1}}}}}\,,
\end{equation}

\begin{equation}
    \alpha_x(\ell)=\ell+\psi_{S_{x_1}}-1\,,
\end{equation}

\noindent and

\begin{equation}
    \beta=e_C+\psi_{S_2}\,,
\end{equation}

\noindent with $\psi_{S_2}\equiv\psi_{S_{x_2}}=1$, and $\ell=i,j$. Now, we need to transform Equation (\ref{eq:joint}) to our ON/OFF jet-axis ratio $R$, Equation (\ref{eq:expected}). Again, leveraging the work of \citet{park2006}, by using their transform

\begin{eqnarray}
    P(\lambda_{\rm S},\lambda_{\rm H}|S,H,B_{\rm S},B_{\rm H}) \, d\lambda_{\rm S} \, d\lambda_{\rm H} &=& P(\lambda_{\rm S}(\mathcal{HR},\omega^\prime),\lambda_{\rm H}(\mathcal{HR},\omega^\prime)|S,H,B_{\rm S},B_{\rm H})\!\left|\frac{\partial(\lambda_{\rm S},\lambda_{\rm H})}{\partial(\mathcal{HR},\omega^\prime)}\right|\!d(\mathcal{HR})d\omega^\prime
    \nonumber\\
    &=&P\!\left(\frac{(1-\mathcal{HR})\omega^\prime}{2},\frac{(1+\mathcal{HR})\omega^\prime}{2}\biggl|S,H,B_{\rm S},B_{\rm H}\!\right)\!\frac{\omega^\prime}{2}d(\mathcal{HR})d\omega^\prime\;\;\;\;\;\;\;\;\;
\end{eqnarray}

\noindent where

\begin{equation}
    \omega^\prime = \lambda_{\rm H}+\lambda_{\rm S}\,,
\end{equation}

\noindent we then obtain

\begin{equation}
    P(\lambda_{\rm ON},\lambda_{\rm OFF}|S_{\rm ON},S_{\rm OFF},B) \, d\lambda_{\rm ON} \, d\lambda_{\rm OFF}=P\!\left(\!\left(1+\frac{R}{2}\right)\!\omega,\left(1-\frac{R}{2}\right)\!\omega\,\biggl|S_{\rm ON},S_{\rm OFF},B\!\right)\!\omega\,dR\,d\omega\,,
\end{equation}

\noindent where
\begin{equation}
    \omega=\frac{\lambda_{\rm ON}+\lambda_{\rm OFF}}{2}\,.
\end{equation}

\noindent So integrating out $\omega$, we have

\begin{eqnarray}
    P(R|S_{\rm ON},S_{\rm OFF},B)&=&\sum_{i=0}^{S_{\rm ON}}\sum_{j=0}^{S_{\rm OFF}}\Lambda_{ij}(S_{\rm ON},S_{\rm OFF},B)\!\!\int_0^\infty\left[\left(1+\frac{R}{2}\right)\!\omega\right]^{\alpha_{\rm ON}(i)}\left[\left(1-\frac{R}{2}\right)\!\omega\right]^{\alpha_{\rm OFF}(j)}
    \nonumber\\&&\times
    \;e\raisebox{3.5ex}{$\;\;\displaystyle-\beta\left[\left(1+\frac{R}{2}\right)\omega+\left(1-\frac{R}{2}\right)\omega\right]$}\;\omega\, d\omega
    \nonumber\\
    &=&\sum_{i=0}^{S_{\rm ON}}\sum_{j=0}^{S_{\rm OFF}}\Lambda_{ij}(S_{\rm ON},S_{\rm OFF},B)\left(1+\frac{R}{2}\right)^{\alpha_{\rm ON}(i)}\left(1-\frac{R}{2}\right)^{\alpha_{\rm OFF}(j)}
    \int_0^\infty\omega^{\alpha_{\rm ON}(i)+\alpha_{\rm OFF}(j)+1}e^{-2\beta\omega}d\omega
    \nonumber\\
    &=&\sum_{i=0}^{S_{\rm ON}}\sum_{j=0}^{S_{\rm OFF}}\kappa_{ij}(S_{\rm ON},S_{\rm OFF},B,R)\int_0^\infty\omega^{\eta_{ij}-1}e^{-2\beta\omega}d\omega\,,
\end{eqnarray}

\noindent where

\begin{equation}
    \Lambda_{ij}(S_{\rm ON},S_{\rm OFF},B)=N(i,S_{\rm ON},B)N(j,S_{\rm OFF})\,,
\end{equation}

\begin{equation}
    \kappa_{ij}(S_{\rm ON},S_{\rm OFF},B,R)=\Lambda_{ij}(S_{\rm ON},S_{\rm OFF},B)\left(1+\frac{R}{2}\right)^{\alpha_{\rm ON}(i)}\left(1-\frac{R}{2}\right)^{\alpha_{\rm OFF}(j)}\,,
\end{equation}

\noindent and

\begin{equation}
    \eta_{ij}=\alpha_{\rm ON}(i)+\alpha_{\rm OFF}(j)+2\,,
\end{equation}

\noindent leaving us with $\int_0^\infty\omega^{\eta_{ij}-1}e^{-2\beta\omega}d\omega$ to evaluate. So,

\begin{eqnarray}
    \int_0^\infty\omega^{\eta_{ij}-1}e^{-2\beta\omega}d\omega&=&\int_0^\infty\frac{(2\beta\omega)^{\eta_{ij}-1}}{(2\beta)^{\eta_{ij}-1\;\;\;\;}}\,e^{-(2\beta\omega)}\frac{d(2\beta\omega)}{2\beta}
    =\frac{1}{(2\beta)^{\eta_{ij}}}\int_0^\infty x^{\eta_{ij}-1}e^{-x}dx
    =\frac{1}{(2\beta)^{\eta_{ij}}}\Gamma(\eta_{ij})
    \nonumber\\
    &=&\frac{\Gamma(\alpha_{\rm ON}(i)+\alpha_{\rm OFF}(j)+2)}{(2\beta)^{\alpha_{\rm ON}(i)+\alpha_{\rm OFF}(j)+2}}
    =\frac{\Gamma(i+j+\psi_{S_{{\rm ON}_1}}+\psi_{S_{{\rm OFF}_1}})}{[2(e_C+\psi_{S_2})]^{i+j+\psi_{S_{{\rm ON}_1}}+\psi_{S_{{\rm OFF}_1}}}}\,.
\end{eqnarray}

\noindent Therefore,

\begin{equation}
    P(R|S_{\rm ON},S_{\rm OFF},B)=\sum_{i=0}^{S_{\rm ON}}\sum_{j=0}^{S_{\rm OFF}}\kappa_{ij}(S_{\rm ON},S_{\rm OFF},B,R)\frac{\Gamma(i+j+\psi_{S_{{\rm ON}_1}}+\psi_{S_{{\rm OFF}_1}})}{[2(e_C+\psi_{S_2})]^{i+j+\psi_{S_{{\rm ON}_1}}+\psi_{S_{{\rm OFF}_1}}}}\,,
\end{equation}

\noindent where

\begin{equation}
    \kappa_{ij}(S_{\rm ON},S_{\rm OFF},B,R)=N(i,S_{\rm ON},B)N(j,S_{\rm OFF},B)\left(1+\frac{R}{2}\right)^{i+\psi_{S_{{\rm ON}_1}}-1}\left(1-\frac{R}{2}\right)^{j+\psi_{S_{{\rm OFF}_1}}-1}\,,
\end{equation}

\noindent and

\begin{equation}
    N(\ell,S_x,B)=\frac{\displaystyle\frac{\Gamma(S_x-\ell+B+\psi_{S_{x_3}})}{\Gamma(\ell+1)\Gamma(S_x-\ell+1)(e_C+r\,e_C+\psi_{S_{x_4}})^{S_x-\ell+B+\psi_{S_{x_3}}}}}{\displaystyle \sum_{k=0}^{S_x}\frac{\Gamma(S_x-k+B+\psi_{S_{x_3}})\Gamma(k+\psi_{S_{x_1}})}{\Gamma(k+1)\Gamma(S_x-k+1)(e_C+r\,e_C+\psi_{S_{x_4}})^{S_x-k+B+\psi_{S_{x_3}}}(e_C+\psi_{S_{x_2}})^{k+\psi_{S_{x_1}}}}}\,.
\end{equation}

\noindent Finally, substituting $r=2$, $e_C=1$, and Equations (\ref{eq:psis}) into above, we obtain Equation (\ref{eq:pr}). Figures~(\ref{fig:r_3c401_broad}) and~(\ref{fig:r_3c22_soft}) show examples of the posterior distributions for $R^{\rm 3C401}_{\rm broad}$ and $R^{\rm 3C22}_{\rm soft}$, respectively, which can be contrasted with Figure~(\ref{fig:diagnostic_plots}).

{Coding up Equation (\ref{eq:pr}) is pretty straightforward, except one needs to pay careful attention to the $\Gamma$ functions and powers, so as to avoid rounding errors. It is recommended one use an arbitrary precision library, such as, \texttt{mpmath}.\footnote{{\texttt{https://mpmath.org}}} Herein, we also use the Gaussian approximation, Equation (\ref{eq:r_approx}), for counts $S_{\rm{ON}}+S_{\rm{OFF}}>2000$. Figure~\ref{fig:r_table} and Table~\ref{tb:r_stats} summarize the results.}
\section{$R$ values} 
\label{ap:r_values}
Table \ref{tb:r_stats} shows the $R$ values for the sources in our sample for the different \textit{Chandra} energy bands. 

\begin{longtable}{lclcccc}
      \caption{$R$ values {(\textit{i.e.}, modes)} for soft, medium, hard, and broad bands, computed using Equations (\ref{eq:pr}) and~(\ref{eq:r_approx})$^\dagger$ with input counts from Table~\ref{tb:on_off_b_counts}. {The error limits are given by the 68.27\% HPD bounds for Equation (\ref{eq:pr}) and by the $\pm1\sigma$ bounds for Equation (\ref{eq:r_approx}).}\label{tb:r_stats}}\\
      \hline\hline
      Source & $z$ & FR type & $R_{\rm{soft}}$ & $R_{\rm{medium}}$ & $R_{\rm{hard}}$ & $R_{\rm{broad}}$ \\
      \hline
      \endfirsthead
      \caption*{$(Continued\mbox{\ldots})$}\\
      \hline\hline
        Source & $z$ & FR type & $R_{\rm{soft}}$ & $R_{\rm{medium}}$ & $R_{\rm{hard}}$ & $R_{\rm{broad}}$ \\
      \hline
      \endhead
      3C 16 & 0.4050 & FR II & $\def\arraystretch{0.6}+0.95\!\begin{array}{l}\mbox{\textbf{$+0.24$}}\\\mbox{\textbf{$-0.27$}}\end{array}$ & $\def\arraystretch{0.6}-0.11\!\begin{array}{l}\mbox{\textbf{$+0.31$}}\\\mbox{\textbf{$-0.29$}}\end{array}$ & $\def\arraystretch{0.6}-0.11\!\begin{array}{l}\mbox{\textbf{$+0.19$}}\\\mbox{\textbf{$-0.18$}}\end{array}$ & $\def\arraystretch{0.6}+0.24\!\begin{array}{l}\mbox{\textbf{$+0.14$}}\\\mbox{\textbf{$-0.14$}}\end{array}$  \\
      3C 19 & 0.4820 & FR II & $\def\arraystretch{0.6}-0.24\!\begin{array}{l}\mbox{\textbf{$+0.29$}}\\\mbox{\textbf{$-0.29$}}\end{array}$ & $\def\arraystretch{0.6}+0.43\!\begin{array}{l}\mbox{\textbf{$+0.32$}}\\\mbox{\textbf{$-0.35$}}\end{array}$ & $\def\arraystretch{0.6}-0.30\!\begin{array}{l}\mbox{\textbf{$+0.60$}}\\\mbox{\textbf{$-0.55$}}\end{array}$ & $\def\arraystretch{0.6}+0.08\!\begin{array}{l}\mbox{\textbf{$+0.22$}}\\\mbox{\textbf{$-0.21$}}\end{array}$  \\
      3C 20 & 0.1740 & FR II & $\def\arraystretch{0.6}-0.07\!\begin{array}{l}\mbox{\textbf{$+0.33$}}\\\mbox{\textbf{$-0.33$}}\end{array}$ & $\def\arraystretch{0.6}+0.51\!\begin{array}{l}\mbox{\textbf{$+0.41$}}\\\mbox{\textbf{$-0.41$}}\end{array}$ & $\def\arraystretch{0.6}+0.04\!\begin{array}{l}\mbox{\textbf{$+0.25$}}\\\mbox{\textbf{$-0.24$}}\end{array}$ & $\def\arraystretch{0.6}+0.07\!\begin{array}{l}\mbox{\textbf{$+0.17$}}\\\mbox{\textbf{$-0.17$}}\end{array}$  \\
      3C 22 & 0.9360 & FR II & $\def\arraystretch{0.6}+1.43\!\begin{array}{l}\mbox{\textbf{$+0.29$}}\\\mbox{\textbf{$-0.37$}}\end{array}$ & $\def\arraystretch{0.6}+1.10\!\begin{array}{l}\mbox{\textbf{$+0.39$}}\\\mbox{\textbf{$-0.43$}}\end{array}$ & $\def\arraystretch{0.6}-0.46\!\begin{array}{l}\mbox{\textbf{$+0.33$}}\\\mbox{\textbf{$-0.30$}}\end{array}$ & $\def\arraystretch{0.6}+0.83\!\begin{array}{l}\mbox{\textbf{$+0.23$}}\\\mbox{\textbf{$-0.21$}}\end{array}$  \\
      3C 29 & 0.0450 & FR I  & $\def\arraystretch{0.6}+0.76\!\begin{array}{l}\mbox{\textbf{$+0.14$}}\\\mbox{\textbf{$-0.14$}}\end{array}$ & $\def\arraystretch{0.6}-0.05\!\begin{array}{l}\mbox{\textbf{$+0.16$}}\\\mbox{\textbf{$-0.15$}}\end{array}$ & $\def\arraystretch{0.6}+0.00\!\begin{array}{l}\mbox{\textbf{$+0.13$}}\\\mbox{\textbf{$-0.13$}}\end{array}$ & $\def\arraystretch{0.6}+0.22\!\begin{array}{l}\mbox{\textbf{$+0.08$}}\\\mbox{\textbf{$-0.08$}}\end{array}$  \\
      3C 34 & 0.6900 & FR II & $\def\arraystretch{0.6}+0.64\!\begin{array}{l}\mbox{\textbf{$+0.34$}}\\\mbox{\textbf{$-0.37$}}\end{array}$ & $\def\arraystretch{0.6}-0.22\!\begin{array}{l}\mbox{\textbf{$+0.30$}}\\\mbox{\textbf{$-0.28$}}\end{array}$ & $\def\arraystretch{0.6}-0.58\!\begin{array}{l}\mbox{\textbf{$+0.27$}}\\\mbox{\textbf{$-0.25$}}\end{array}$ & $\def\arraystretch{0.6}+0.43\!\begin{array}{l}\mbox{\textbf{$+0.17$}}\\\mbox{\textbf{$-0.17$}}\end{array}$  \\
      3C 42 & 0.3950 & FR II & $\def\arraystretch{0.6}+1.70\!\begin{array}{l}\mbox{\textbf{$+0.25$}}\\\mbox{\textbf{$-0.38$}}\end{array}$ & $\def\arraystretch{0.6}-1.00\!\begin{array}{l}\mbox{\textbf{$+0.70$}}\\\mbox{\textbf{$-0.50$}}\end{array}$ & $\def\arraystretch{0.6}-0.11\!\begin{array}{l}\mbox{\textbf{$+0.39$}}\\\mbox{\textbf{$-0.34$}}\end{array}$ & $\def\arraystretch{0.6}+0.33\!\begin{array}{l}\mbox{\textbf{$+0.27$}}\\\mbox{\textbf{$-0.28$}}\end{array}$  \\
      3C 46 & 0.4373 & FR II & $\def\arraystretch{0.6}-0.03\!\begin{array}{l}\mbox{\textbf{$+0.15$}}\\\mbox{\textbf{$-0.14$}}\end{array}$ & $\def\arraystretch{0.6}-0.22\!\begin{array}{l}\mbox{\textbf{$+0.16$}}\\\mbox{\textbf{$-0.15$}}\end{array}$ & $\def\arraystretch{0.6}+0.08\!\begin{array}{l}\mbox{\textbf{$+0.09$}}\\\mbox{\textbf{$-0.09$}}\end{array}$ & $\def\arraystretch{0.6}-0.08\!\begin{array}{l}\mbox{\textbf{$+0.07$}}\\\mbox{\textbf{$-0.07$}}\end{array}$  \\
      3C 52 & 0.2900 & FR II & $\def\arraystretch{0.6}+0.55\!\begin{array}{l}\mbox{\textbf{$+0.21$}}\\\mbox{\textbf{$-0.21$}}\end{array}$ & $\def\arraystretch{0.6}+0.13\!\begin{array}{l}\mbox{\textbf{$+0.17$}}\\\mbox{\textbf{$-0.18$}}\end{array}$ & $\def\arraystretch{0.6}-0.08\!\begin{array}{l}\mbox{\textbf{$+0.16$}}\\\mbox{\textbf{$-0.16$}}\end{array}$ & $\def\arraystretch{0.6}+0.19\!\begin{array}{l}\mbox{\textbf{$+0.10$}}\\\mbox{\textbf{$-0.10$}}\end{array}$  \\
      3C 54 & 0.8274 & FR II & $\def\arraystretch{0.6}+0.14\!\begin{array}{l}\mbox{\textbf{$+0.41$}}\\\mbox{\textbf{$-0.41$}}\end{array}$ & $\def\arraystretch{0.6}+0.93\!\begin{array}{l}\mbox{\textbf{$+0.33$}}\\\mbox{\textbf{$-0.33$}}\end{array}$ & $\def\arraystretch{0.6}-0.24\!\begin{array}{l}\mbox{\textbf{$+0.20$}}\\\mbox{\textbf{$-0.19$}}\end{array}$ & $\def\arraystretch{0.6}+0.41\!\begin{array}{l}\mbox{\textbf{$+0.16$}}\\\mbox{\textbf{$-0.15$}}\end{array}$  \\
      3C 55 & 0.7348 & FR II & $\def\arraystretch{0.6}+1.24\!\begin{array}{l}\mbox{\textbf{$+0.23$}}\\\mbox{\textbf{$-0.25$}}\end{array}$ & $\def\arraystretch{0.6}+1.18\!\begin{array}{l}\mbox{\textbf{$+0.24$}}\\\mbox{\textbf{$-0.26$}}\end{array}$ & $\def\arraystretch{0.6}-0.22\!\begin{array}{l}\mbox{\textbf{$+0.23$}}\\\mbox{\textbf{$-0.20$}}\end{array}$ & $\def\arraystretch{0.6}+0.66\!\begin{array}{l}\mbox{\textbf{$+0.14$}}\\\mbox{\textbf{$-0.15$}}\end{array}$  \\
      3C 79 & 0.2559 & FR II & $\def\arraystretch{0.6}+0.24\!\begin{array}{l}\mbox{\textbf{$+0.25$}}\\\mbox{\textbf{$-0.25$}}\end{array}$ & $\def\arraystretch{0.6}+0.10\!\begin{array}{l}\mbox{\textbf{$+0.26$}}\\\mbox{\textbf{$-0.26$}}\end{array}$ & $\def\arraystretch{0.6}-0.13\!\begin{array}{l}\mbox{\textbf{$+0.19$}}\\\mbox{\textbf{$-0.18$}}\end{array}$ & $\def\arraystretch{0.6}+0.17\!\begin{array}{l}\mbox{\textbf{$+0.13$}}\\\mbox{\textbf{$-0.12$}}\end{array}$  \\
      3C 129.1 & 0.0222 & FR I  & $\def\arraystretch{0.6}+0.01\!\begin{array}{l}\mbox{\textbf{$+0.15$}}\\\mbox{\textbf{$-0.14$}}\end{array}$ & $\def\arraystretch{0.6}+0.01\!\begin{array}{l}\mbox{\textbf{$+0.08$}}\\\mbox{\textbf{$-0.08$}}\end{array}$ & $\def\arraystretch{0.6}-0.17\!\begin{array}{l}\mbox{\textbf{$+0.06$}}\\\mbox{\textbf{$-0.06$}}\end{array}$ & $\def\arraystretch{0.6}+0.10\!\begin{array}{l}\mbox{\textbf{$+0.05$}}\\\mbox{\textbf{$-0.05$}}\end{array}$  \\
      3C 133 & 0.2775 & FR II & $\def\arraystretch{0.6}+0.18\!\begin{array}{l}\mbox{\textbf{$+0.50$}}\\\mbox{\textbf{$-0.45$}}\end{array}$ & $\def\arraystretch{0.6}-0.12\!\begin{array}{l}\mbox{\textbf{$+0.41$}}\\\mbox{\textbf{$-0.41$}}\end{array}$ & $\def\arraystretch{0.6}-0.66\!\begin{array}{l}\mbox{\textbf{$+0.29$}}\\\mbox{\textbf{$-0.26$}}\end{array}$ & $\def\arraystretch{0.6}+0.38\!\begin{array}{l}\mbox{\textbf{$+0.21$}}\\\mbox{\textbf{$-0.19$}}\end{array}$  \\
      3C 135 & 0.1273 & FR II & $\def\arraystretch{0.6}+0.02\!\begin{array}{l}\mbox{\textbf{$+0.19$}}\\\mbox{\textbf{$-0.18$}}\end{array}$ & $\def\arraystretch{0.6}+0.18\!\begin{array}{l}\mbox{\textbf{$+0.21$}}\\\mbox{\textbf{$-0.20$}}\end{array}$ & $\def\arraystretch{0.6}+0.11\!\begin{array}{l}\mbox{\textbf{$+0.12$}}\\\mbox{\textbf{$-0.11$}}\end{array}$ & $\def\arraystretch{0.6}-0.03\!\begin{array}{l}\mbox{\textbf{$+0.09$}}\\\mbox{\textbf{$-0.09$}}\end{array}$  \\
      3C 166 & 0.2449 & FR II & $\def\arraystretch{0.6}+0.68\!\begin{array}{l}\mbox{\textbf{$+0.28$}}\\\mbox{\textbf{$-0.28$}}\end{array}$ & $\def\arraystretch{0.6}+0.70\!\begin{array}{l}\mbox{\textbf{$+0.25$}}\\\mbox{\textbf{$-0.25$}}\end{array}$ & $\def\arraystretch{0.6}-0.38\!\begin{array}{l}\mbox{\textbf{$+0.33$}}\\\mbox{\textbf{$-0.29$}}\end{array}$ & $\def\arraystretch{0.6}+0.58\!\begin{array}{l}\mbox{\textbf{$+0.17$}}\\\mbox{\textbf{$-0.16$}}\end{array}$  \\
      3C 169.1 & 0.6330 & FR II & $\def\arraystretch{0.6}+1.06\!\begin{array}{l}\mbox{\textbf{$+0.30$}}\\\mbox{\textbf{$-0.30$}}\end{array}$ & $\def\arraystretch{0.6}+1.03\!\begin{array}{l}\mbox{\textbf{$+0.33$}}\\\mbox{\textbf{$-0.37$}}\end{array}$ & $\def\arraystretch{0.6}-0.20\!\begin{array}{l}\mbox{\textbf{$+0.28$}}\\\mbox{\textbf{$-0.27$}}\end{array}$ & $\def\arraystretch{0.6}+0.61\!\begin{array}{l}\mbox{\textbf{$+0.17$}}\\\mbox{\textbf{$-0.19$}}\end{array}$  \\
      3C 171 & 0.2384 & FR II & $\def\arraystretch{0.6}+0.17\!\begin{array}{l}\mbox{\textbf{$+0.17$}}\\\mbox{\textbf{$-0.16$}}\end{array}$ & $\def\arraystretch{0.6}+0.34\!\begin{array}{l}\mbox{\textbf{$+0.20$}}\\\mbox{\textbf{$-0.19$}}\end{array}$ & $\def\arraystretch{0.6}+0.13\!\begin{array}{l}\mbox{\textbf{$+0.12$}}\\\mbox{\textbf{$-0.12$}}\end{array}$ & $\def\arraystretch{0.6}+0.05\!\begin{array}{l}\mbox{\textbf{$+0.09$}}\\\mbox{\textbf{$-0.08$}}\end{array}$  \\
      3C 175.1 & 0.9200 & FR II & $\def\arraystretch{0.6}-0.70\!\begin{array}{l}\mbox{\textbf{$+0.60$}}\\\mbox{\textbf{$-0.50$}}\end{array}$ & $\def\arraystretch{0.6}+0.60\!\begin{array}{l}\mbox{\textbf{$+0.42$}}\\\mbox{\textbf{$-0.47$}}\end{array}$ & $\def\arraystretch{0.6}+0.00\!\begin{array}{l}\mbox{\textbf{$+0.47$}}\\\mbox{\textbf{$-0.42$}}\end{array}$ & $\def\arraystretch{0.6}+0.14\!\begin{array}{l}\mbox{\textbf{$+0.32$}}\\\mbox{\textbf{$-0.30$}}\end{array}$  \\
      3C 184.1 & 0.1182 & FR II & $\def\arraystretch{0.6}+0.26\!\begin{array}{l}\mbox{\textbf{$+0.16$}}\\\mbox{\textbf{$-0.15$}}\end{array}$ & $\def\arraystretch{0.6}+0.27\!\begin{array}{l}\mbox{\textbf{$+0.17$}}\\\mbox{\textbf{$-0.16$}}\end{array}$ & $\def\arraystretch{0.6}+0.04\!\begin{array}{l}\mbox{\textbf{$+0.09$}}\\\mbox{\textbf{$-0.09$}}\end{array}$ & $\def\arraystretch{0.6}+0.08\!\begin{array}{l}\mbox{\textbf{$+0.07$}}\\\mbox{\textbf{$-0.07$}}\end{array}$  \\
      3C 192 & 0.0597 & FR II & $\def\arraystretch{0.6}+0.46\!\begin{array}{l}\mbox{\textbf{$+0.11$}}\\\mbox{\textbf{$-0.11$}}\end{array}$ & $\def\arraystretch{0.6}+0.21\!\begin{array}{l}\mbox{\textbf{$+0.13$}}\\\mbox{\textbf{$-0.12$}}\end{array}$ & $\def\arraystretch{0.6}-0.11\!\begin{array}{l}\mbox{\textbf{$+0.08$}}\\\mbox{\textbf{$-0.08$}}\end{array}$ & $\def\arraystretch{0.6}+0.22\!\begin{array}{l}\mbox{\textbf{$+0.06$}}\\\mbox{\textbf{$-0.06$}}\end{array}$  \\
      3C 215 & 0.4121 & FR I  & $\def\arraystretch{0.6}+0.03\!\begin{array}{l}\mbox{\textbf{$+0.08$}}\\\mbox{\textbf{$-0.07$}}\end{array}$ & $\def\arraystretch{0.6}+0.29\!\begin{array}{l}\mbox{\textbf{$+0.10$}}\\\mbox{\textbf{$-0.09$}}\end{array}$ & $\def\arraystretch{0.6}-0.24\!\begin{array}{l}\mbox{\textbf{$+0.09$}}\\\mbox{\textbf{$-0.09$}}\end{array}$ & $\def\arraystretch{0.6}+0.16\!\begin{array}{l}\mbox{\textbf{$+0.05$}}\\\mbox{\textbf{$-0.05$}}\end{array}$  \\
      3C 223.1 & 0.1075 & FR II & $\def\arraystretch{0.6}+0.29\!\begin{array}{l}\mbox{\textbf{$+0.35$}}\\\mbox{\textbf{$-0.37$}}\end{array}$ & $\def\arraystretch{0.6}+0.52\!\begin{array}{l}\mbox{\textbf{$+0.32$}}\\\mbox{\textbf{$-0.32$}}\end{array}$ & $\def\arraystretch{0.6}-0.17\!\begin{array}{l}\mbox{\textbf{$+0.22$}}\\\mbox{\textbf{$-0.20$}}\end{array}$ & $\def\arraystretch{0.6}+0.32\!\begin{array}{l}\mbox{\textbf{$+0.16$}}\\\mbox{\textbf{$-0.15$}}\end{array}$  \\
      3C 234 & 0.1849 & FR II & $\def\arraystretch{0.6}+0.16\!\begin{array}{l}\mbox{\textbf{$+0.17$}}\\\mbox{\textbf{$-0.17$}}\end{array}$ & $\def\arraystretch{0.6}+0.19\!\begin{array}{l}\mbox{\textbf{$+0.22$}}\\\mbox{\textbf{$-0.21$}}\end{array}$ & $\def\arraystretch{0.6}-0.14\!\begin{array}{l}\mbox{\textbf{$+0.13$}}\\\mbox{\textbf{$-0.12$}}\end{array}$ & $\def\arraystretch{0.6}+0.19\!\begin{array}{l}\mbox{\textbf{$+0.10$}}\\\mbox{\textbf{$-0.09$}}\end{array}$  \\
      3C 244.1 & 0.4280 & FR II & $\def\arraystretch{0.6}+0.48\!\begin{array}{l}\mbox{\textbf{$+0.26$}}\\\mbox{\textbf{$-0.26$}}\end{array}$ & $\def\arraystretch{0.6}+0.30\!\begin{array}{l}\mbox{\textbf{$+0.26$}}\\\mbox{\textbf{$-0.28$}}\end{array}$ & $\def\arraystretch{0.6}-0.45\!\begin{array}{l}\mbox{\textbf{$+0.22$}}\\\mbox{\textbf{$-0.22$}}\end{array}$ & $\def\arraystretch{0.6}+0.37\!\begin{array}{l}\mbox{\textbf{$+0.16$}}\\\mbox{\textbf{$-0.15$}}\end{array}$  \\
      3C 268.2 & 0.3620 & FR II & $\def\arraystretch{0.6}+0.29\!\begin{array}{l}\mbox{\textbf{$+0.24$}}\\\mbox{\textbf{$-0.26$}}\end{array}$ & $\def\arraystretch{0.6}+0.45\!\begin{array}{l}\mbox{\textbf{$+0.24$}}\\\mbox{\textbf{$-0.22$}}\end{array}$ & $\def\arraystretch{0.6}+0.10\!\begin{array}{l}\mbox{\textbf{$+0.17$}}\\\mbox{\textbf{$-0.18$}}\end{array}$ & $\def\arraystretch{0.6}+0.13\!\begin{array}{l}\mbox{\textbf{$+0.12$}}\\\mbox{\textbf{$-0.12$}}\end{array}$  \\
      3C 272 & 0.9440 & FR II & $\def\arraystretch{0.6}+0.98\!\begin{array}{l}\mbox{\textbf{$+0.35$}}\\\mbox{\textbf{$-0.35$}}\end{array}$ & $\def\arraystretch{0.6}+0.20\!\begin{array}{l}\mbox{\textbf{$+0.31$}}\\\mbox{\textbf{$-0.29$}}\end{array}$ & $\def\arraystretch{0.6}-0.17\!\begin{array}{l}\mbox{\textbf{$+0.21$}}\\\mbox{\textbf{$-0.21$}}\end{array}$ & $\def\arraystretch{0.6}+0.31\!\begin{array}{l}\mbox{\textbf{$+0.16$}}\\\mbox{\textbf{$-0.16$}}\end{array}$  \\
      3C 277 & 0.4140 & FR I  & $\def\arraystretch{0.6}+0.28\!\begin{array}{l}\mbox{\textbf{$+0.18$}}\\\mbox{\textbf{$-0.17$}}\end{array}$ & $\def\arraystretch{0.6}+0.09\!\begin{array}{l}\mbox{\textbf{$+0.17$}}\\\mbox{\textbf{$-0.16$}}\end{array}$ & $\def\arraystretch{0.6}-0.22\!\begin{array}{l}\mbox{\textbf{$+0.11$}}\\\mbox{\textbf{$-0.10$}}\end{array}$ & $\def\arraystretch{0.6}+0.21\!\begin{array}{l}\mbox{\textbf{$+0.08$}}\\\mbox{\textbf{$-0.08$}}\end{array}$  \\
      3C 284 & 0.2398 & FR II & $\def\arraystretch{0.6}+0.34\!\begin{array}{l}\mbox{\textbf{$+0.14$}}\\\mbox{\textbf{$-0.13$}}\end{array}$ & $\def\arraystretch{0.6}-0.21\!\begin{array}{l}\mbox{\textbf{$+0.14$}}\\\mbox{\textbf{$-0.14$}}\end{array}$ & $\def\arraystretch{0.6}-0.15\!\begin{array}{l}\mbox{\textbf{$+0.08$}}\\\mbox{\textbf{$-0.08$}}\end{array}$ & $\def\arraystretch{0.6}+0.20\!\begin{array}{l}\mbox{\textbf{$+0.07$}}\\\mbox{\textbf{$-0.06$}}\end{array}$  \\
      3C 285 & 0.0794 & FR II & $\def\arraystretch{0.6}-0.01\!\begin{array}{l}\mbox{\textbf{$+0.06$}}\\\mbox{\textbf{$-0.06$}}\end{array}$ & $\def\arraystretch{0.6}-0.07\!\begin{array}{l}\mbox{\textbf{$+0.07$}}\\\mbox{\textbf{$-0.07$}}\end{array}$ & $\def\arraystretch{0.6}+0.02\!\begin{array}{l}\mbox{\textbf{$+0.05$}}\\\mbox{\textbf{$-0.05$}}\end{array}$ & $-0.10\mbox{\textbf{$\pm0.19$}}^\dagger$  \\
      3C 287.1 & 0.2156 & FR II & $\def\arraystretch{0.6}+0.39\!\begin{array}{l}\mbox{\textbf{$+0.11$}}\\\mbox{\textbf{$-0.12$}}\end{array}$ & $\def\arraystretch{0.6}+0.52\!\begin{array}{l}\mbox{\textbf{$+0.14$}}\\\mbox{\textbf{$-0.13$}}\end{array}$ & $\def\arraystretch{0.6}-0.15\!\begin{array}{l}\mbox{\textbf{$+0.11$}}\\\mbox{\textbf{$-0.11$}}\end{array}$ & $\def\arraystretch{0.6}+0.30\!\begin{array}{l}\mbox{\textbf{$+0.06$}}\\\mbox{\textbf{$-0.07$}}\end{array}$  \\
      3C 289 & 0.9674 & FR II & $\def\arraystretch{0.6}+0.70\!\begin{array}{l}\mbox{\textbf{$+0.70$}}\\\mbox{\textbf{$-0.80$}}\end{array}$ & $\def\arraystretch{0.6}+0.42\!\begin{array}{l}\mbox{\textbf{$+0.50$}}\\\mbox{\textbf{$-0.50$}}\end{array}$ & $\def\arraystretch{0.6}-0.61\!\begin{array}{l}\mbox{\textbf{$+0.46$}}\\\mbox{\textbf{$-0.39$}}\end{array}$ & $\def\arraystretch{0.6}+0.64\!\begin{array}{l}\mbox{\textbf{$+0.28$}}\\\mbox{\textbf{$-0.28$}}\end{array}$  \\
      3C 292 & 0.7100 & FR II & $\def\arraystretch{0.6}+0.44\!\begin{array}{l}\mbox{\textbf{$+0.16$}}\\\mbox{\textbf{$-0.16$}}\end{array}$ & $\def\arraystretch{0.6}+0.56\!\begin{array}{l}\mbox{\textbf{$+0.15$}}\\\mbox{\textbf{$-0.14$}}\end{array}$ & $\def\arraystretch{0.6}+0.01\!\begin{array}{l}\mbox{\textbf{$+0.10$}}\\\mbox{\textbf{$-0.10$}}\end{array}$ & $\def\arraystretch{0.6}+0.21\!\begin{array}{l}\mbox{\textbf{$+0.07$}}\\\mbox{\textbf{$-0.07$}}\end{array}$  \\
      3C 293.1 & 0.7090 & FR II & $\def\arraystretch{0.6}+0.46\!\begin{array}{l}\mbox{\textbf{$+0.26$}}\\\mbox{\textbf{$-0.28$}}\end{array}$ & $\def\arraystretch{0.6}+0.83\!\begin{array}{l}\mbox{\textbf{$+0.26$}}\\\mbox{\textbf{$-0.26$}}\end{array}$ & $\def\arraystretch{0.6}+0.11\!\begin{array}{l}\mbox{\textbf{$+0.22$}}\\\mbox{\textbf{$-0.21$}}\end{array}$ & $\def\arraystretch{0.6}+0.31\!\begin{array}{l}\mbox{\textbf{$+0.14$}}\\\mbox{\textbf{$-0.15$}}\end{array}$  \\
      3C 300 & 0.2700 & FR II & $\def\arraystretch{0.6}+0.42\!\begin{array}{l}\mbox{\textbf{$+0.20$}}\\\mbox{\textbf{$-0.20$}}\end{array}$ & $\def\arraystretch{0.6}+0.62\!\begin{array}{l}\mbox{\textbf{$+0.22$}}\\\mbox{\textbf{$-0.22$}}\end{array}$ & $\def\arraystretch{0.6}+0.01\!\begin{array}{l}\mbox{\textbf{$+0.15$}}\\\mbox{\textbf{$-0.14$}}\end{array}$ & $\def\arraystretch{0.6}+0.23\!\begin{array}{l}\mbox{\textbf{$+0.11$}}\\\mbox{\textbf{$-0.10$}}\end{array}$  \\
      3C 306.1 & 0.4410 & FR II & $\def\arraystretch{0.6}+0.10\!\begin{array}{l}\mbox{\textbf{$+0.22$}}\\\mbox{\textbf{$-0.21$}}\end{array}$ & $\def\arraystretch{0.6}+0.45\!\begin{array}{l}\mbox{\textbf{$+0.23$}}\\\mbox{\textbf{$-0.23$}}\end{array}$ & $\def\arraystretch{0.6}-0.75\!\begin{array}{l}\mbox{\textbf{$+0.15$}}\\\mbox{\textbf{$-0.14$}}\end{array}$ & $\def\arraystretch{0.6}+0.51\!\begin{array}{l}\mbox{\textbf{$+0.11$}}\\\mbox{\textbf{$-0.11$}}\end{array}$  \\
      3C 315 & 0.1083 & FR I  & $\def\arraystretch{0.6}+0.17\!\begin{array}{l}\mbox{\textbf{$+0.15$}}\\\mbox{\textbf{$-0.15$}}\end{array}$ & $\def\arraystretch{0.6}+0.09\!\begin{array}{l}\mbox{\textbf{$+0.17$}}\\\mbox{\textbf{$-0.16$}}\end{array}$ & $\def\arraystretch{0.6}-0.27\!\begin{array}{l}\mbox{\textbf{$+0.11$}}\\\mbox{\textbf{$-0.10$}}\end{array}$ & $\def\arraystretch{0.6}+0.20\!\begin{array}{l}\mbox{\textbf{$+0.08$}}\\\mbox{\textbf{$-0.08$}}\end{array}$  \\
      3C 319 & 0.1920 & FR II & $\def\arraystretch{0.6}+0.16\!\begin{array}{l}\mbox{\textbf{$+0.20$}}\\\mbox{\textbf{$-0.20$}}\end{array}$ & $\def\arraystretch{0.6}+0.51\!\begin{array}{l}\mbox{\textbf{$+0.23$}}\\\mbox{\textbf{$-0.24$}}\end{array}$ & $\def\arraystretch{0.6}-0.16\!\begin{array}{l}\mbox{\textbf{$+0.17$}}\\\mbox{\textbf{$-0.16$}}\end{array}$ & $\def\arraystretch{0.6}+0.24\!\begin{array}{l}\mbox{\textbf{$+0.11$}}\\\mbox{\textbf{$-0.11$}}\end{array}$  \\
      3C 323.1 & 0.2643 & FR II & $\def\arraystretch{0.6}+0.04\!\begin{array}{l}\mbox{\textbf{$+0.09$}}\\\mbox{\textbf{$-0.08$}}\end{array}$ & $\def\arraystretch{0.6}-0.15\!\begin{array}{l}\mbox{\textbf{$+0.10$}}\\\mbox{\textbf{$-0.10$}}\end{array}$ & $\def\arraystretch{0.6}-0.13\!\begin{array}{l}\mbox{\textbf{$+0.12$}}\\\mbox{\textbf{$-0.12$}}\end{array}$ & $\def\arraystretch{0.6}+0.02\!\begin{array}{l}\mbox{\textbf{$+0.06$}}\\\mbox{\textbf{$-0.05$}}\end{array}$  \\
      3C 327.1 & 0.4620 & FR I  & $\def\arraystretch{0.6}-0.02\!\begin{array}{l}\mbox{\textbf{$+0.14$}}\\\mbox{\textbf{$-0.14$}}\end{array}$ & $\def\arraystretch{0.6}+0.49\!\begin{array}{l}\mbox{\textbf{$+0.17$}}\\\mbox{\textbf{$-0.16$}}\end{array}$ & $\def\arraystretch{0.6}-0.05\!\begin{array}{l}\mbox{\textbf{$+0.18$}}\\\mbox{\textbf{$-0.18$}}\end{array}$ & $\def\arraystretch{0.6}+0.12\!\begin{array}{l}\mbox{\textbf{$+0.09$}}\\\mbox{\textbf{$-0.09$}}\end{array}$  \\
      3C 332 & 0.1510 & FR II & $\def\arraystretch{0.6}+0.26\!\begin{array}{l}\mbox{\textbf{$+0.13$}}\\\mbox{\textbf{$-0.13$}}\end{array}$ & $\def\arraystretch{0.6}+0.55\!\begin{array}{l}\mbox{\textbf{$+0.11$}}\\\mbox{\textbf{$-0.11$}}\end{array}$ & $\def\arraystretch{0.6}+0.09\!\begin{array}{l}\mbox{\textbf{$+0.08$}}\\\mbox{\textbf{$-0.08$}}\end{array}$ & $\def\arraystretch{0.6}+0.14\!\begin{array}{l}\mbox{\textbf{$+0.06$}}\\\mbox{\textbf{$-0.06$}}\end{array}$  \\
      3C 336 & 0.9265 & FR II & $\def\arraystretch{0.6}-2.00\!\begin{array}{l}\mbox{\textbf{$+0.45$}}\\\mbox{\textbf{$-0.00$}}\end{array}$ & $\def\arraystretch{0.6}+2.00\!\begin{array}{l}\mbox{\textbf{$+0.00$}}\\\mbox{\textbf{$-0.35$}}\end{array}$ & $\def\arraystretch{0.6}-1.00\!\begin{array}{l}\mbox{\textbf{$+0.65$}}\\\mbox{\textbf{$-0.45$}}\end{array}$ & $\def\arraystretch{0.6}+0.59\!\begin{array}{l}\mbox{\textbf{$+0.37$}}\\\mbox{\textbf{$-0.37$}}\end{array}$  \\
      3C 340 & 0.7754 & FR II & $\def\arraystretch{0.6}+0.65\!\begin{array}{l}\mbox{\textbf{$+0.43$}}\\\mbox{\textbf{$-0.43$}}\end{array}$ & $\def\arraystretch{0.6}+1.25\!\begin{array}{l}\mbox{\textbf{$+0.29$}}\\\mbox{\textbf{$-0.36$}}\end{array}$ & $\def\arraystretch{0.6}-0.79\!\begin{array}{l}\mbox{\textbf{$+0.31$}}\\\mbox{\textbf{$-0.26$}}\end{array}$ & $\def\arraystretch{0.6}+0.84\!\begin{array}{l}\mbox{\textbf{$+0.20$}}\\\mbox{\textbf{$-0.20$}}\end{array}$  \\
      3C 341 & 0.4480 & FR II & $\def\arraystretch{0.6}-0.05\!\begin{array}{l}\mbox{\textbf{$+0.30$}}\\\mbox{\textbf{$-0.30$}}\end{array}$ & $\def\arraystretch{0.6}+0.73\!\begin{array}{l}\mbox{\textbf{$+0.32$}}\\\mbox{\textbf{$-0.35$}}\end{array}$ & $\def\arraystretch{0.6}-0.40\!\begin{array}{l}\mbox{\textbf{$+0.22$}}\\\mbox{\textbf{$-0.20$}}\end{array}$ & $\def\arraystretch{0.6}+0.34\!\begin{array}{l}\mbox{\textbf{$+0.15$}}\\\mbox{\textbf{$-0.15$}}\end{array}$  \\
      3C 352 & 0.8067 & FR II & $\def\arraystretch{0.6}+1.06\!\begin{array}{l}\mbox{\textbf{$+0.43$}}\\\mbox{\textbf{$-0.47$}}\end{array}$ & $\def\arraystretch{0.6}+0.94\!\begin{array}{l}\mbox{\textbf{$+0.37$}}\\\mbox{\textbf{$-0.41$}}\end{array}$ & $\def\arraystretch{0.6}+1.40\!\begin{array}{l}\mbox{\textbf{$+0.30$}}\\\mbox{\textbf{$-0.42$}}\end{array}$ & $\def\arraystretch{0.6}+0.26\!\begin{array}{l}\mbox{\textbf{$+0.26$}}\\\mbox{\textbf{$-0.26$}}\end{array}$  \\
      3C 357 & 0.1661 & FR II & $\def\arraystretch{0.6}+0.34\!\begin{array}{l}\mbox{\textbf{$+0.24$}}\\\mbox{\textbf{$-0.22$}}\end{array}$ & $\def\arraystretch{0.6}+0.92\!\begin{array}{l}\mbox{\textbf{$+0.21$}}\\\mbox{\textbf{$-0.23$}}\end{array}$ & $\def\arraystretch{0.6}-0.06\!\begin{array}{l}\mbox{\textbf{$+0.18$}}\\\mbox{\textbf{$-0.17$}}\end{array}$ & $\def\arraystretch{0.6}+0.34\!\begin{array}{l}\mbox{\textbf{$+0.12$}}\\\mbox{\textbf{$-0.12$}}\end{array}$  \\
      3C 381 & 0.1605 & FR II & $\def\arraystretch{0.6}+0.83\!\begin{array}{l}\mbox{\textbf{$+0.27$}}\\\mbox{\textbf{$-0.30$}}\end{array}$ & $\def\arraystretch{0.6}+0.70\!\begin{array}{l}\mbox{\textbf{$+0.37$}}\\\mbox{\textbf{$-0.37$}}\end{array}$ & $\def\arraystretch{0.6}-0.15\!\begin{array}{l}\mbox{\textbf{$+0.24$}}\\\mbox{\textbf{$-0.22$}}\end{array}$ & $\def\arraystretch{0.6}+0.43\!\begin{array}{l}\mbox{\textbf{$+0.17$}}\\\mbox{\textbf{$-0.16$}}\end{array}$  \\
      3C 382 & 0.0579 & FR II & $\mbox{\textbf{$+0.14\pm0.06^\dagger$}}$ & $+0.51\mbox{\textbf{$\pm0.06$}}^\dagger$ & $-0.18\mbox{\textbf{$\pm0.06$}}^\dagger$ & $+0.23\mbox{\textbf{$\pm0.05$}}^\dagger$  \\
      3C 388 & 0.0917 & FR II & $\def\arraystretch{0.6}-0.03\!\begin{array}{l}\mbox{\textbf{$+0.04$}}\\\mbox{\textbf{$-0.04$}}\end{array}$ & $\def\arraystretch{0.6}+0.08\!\begin{array}{l}\mbox{\textbf{$+0.04$}}\\\mbox{\textbf{$-0.04$}}\end{array}$ & $\def\arraystretch{0.6}-0.07\!\begin{array}{l}\mbox{\textbf{$+0.05$}}\\\mbox{\textbf{$-0.05$}}\end{array}$ & $+0.04\mbox{\textbf{$\pm0.05$}}^\dagger$  \\
      3C 401 & 0.2011 & FR II & $\def\arraystretch{0.6}-0.25\!\begin{array}{l}\mbox{\textbf{$+0.05$}}\\\mbox{\textbf{$-0.05$}}\end{array}$ & $\def\arraystretch{0.6}-0.08\!\begin{array}{l}\mbox{\textbf{$+0.07$}}\\\mbox{\textbf{$-0.07$}}\end{array}$ & $\def\arraystretch{0.6}+0.15\!\begin{array}{l}\mbox{\textbf{$+0.09$}}\\\mbox{\textbf{$-0.08$}}\end{array}$ & $\def\arraystretch{0.6}-0.17\!\begin{array}{l}\mbox{\textbf{$+0.04$}}\\\mbox{\textbf{$-0.04$}}\end{array}$  \\
      3C 411 & 0.4670 & FR II & $\def\arraystretch{0.6}+0.14\!\begin{array}{l}\mbox{\textbf{$+0.19$}}\\\mbox{\textbf{$-0.17$}}\end{array}$ & $\def\arraystretch{0.6}-0.05\!\begin{array}{l}\mbox{\textbf{$+0.16$}}\\\mbox{\textbf{$-0.16$}}\end{array}$ & $\def\arraystretch{0.6}-0.05\!\begin{array}{l}\mbox{\textbf{$+0.17$}}\\\mbox{\textbf{$-0.17$}}\end{array}$ & $\def\arraystretch{0.6}+0.10\!\begin{array}{l}\mbox{\textbf{$+0.10$}}\\\mbox{\textbf{$-0.10$}}\end{array}$  \\
      3C 424 & 0.1270 & FR I  & $\def\arraystretch{0.6}-0.07\!\begin{array}{l}\mbox{\textbf{$+0.30$}}\\\mbox{\textbf{$-0.30$}}\end{array}$ & $\def\arraystretch{0.6}+0.61\!\begin{array}{l}\mbox{\textbf{$+0.31$}}\\\mbox{\textbf{$-0.34$}}\end{array}$ & $\def\arraystretch{0.6}-0.07\!\begin{array}{l}\mbox{\textbf{$+0.35$}}\\\mbox{\textbf{$-0.35$}}\end{array}$ & $\def\arraystretch{0.6}+0.35\!\begin{array}{l}\mbox{\textbf{$+0.20$}}\\\mbox{\textbf{$-0.20$}}\end{array}$  \\
      3C 427.1 & 0.5720 & FR II & $\def\arraystretch{0.6}+0.14\!\begin{array}{l}\mbox{\textbf{$+0.16$}}\\\mbox{\textbf{$-0.16$}}\end{array}$ & $\def\arraystretch{0.6}+0.39\!\begin{array}{l}\mbox{\textbf{$+0.16$}}\\\mbox{\textbf{$-0.17$}}\end{array}$ & $\def\arraystretch{0.6}-0.49\!\begin{array}{l}\mbox{\textbf{$+0.20$}}\\\mbox{\textbf{$-0.19$}}\end{array}$ & $\def\arraystretch{0.6}+0.33\!\begin{array}{l}\mbox{\textbf{$+0.10$}}\\\mbox{\textbf{$-0.10$}}\end{array}$  \\
      3C 432 & 1.7850 & FR II & $\def\arraystretch{0.6}+0.58\!\begin{array}{l}\mbox{\textbf{$+0.25$}}\\\mbox{\textbf{$-0.25$}}\end{array}$ & $\def\arraystretch{0.6}+0.13\!\begin{array}{l}\mbox{\textbf{$+0.28$}}\\\mbox{\textbf{$-0.26$}}\end{array}$ & $\def\arraystretch{0.6}+0.00\!\begin{array}{l}\mbox{\textbf{$+0.28$}}\\\mbox{\textbf{$-0.26$}}\end{array}$ & $\def\arraystretch{0.6}+0.37\!\begin{array}{l}\mbox{\textbf{$+0.15$}}\\\mbox{\textbf{$-0.14$}}\end{array}$  \\
      3C 434 & 0.3220 & FR II & $\def\arraystretch{0.6}-1.24\!\begin{array}{l}\mbox{\textbf{$+0.50$}}\\\mbox{\textbf{$-0.37$}}\end{array}$ & $\def\arraystretch{0.6}+0.45\!\begin{array}{l}\mbox{\textbf{$+0.65$}}\\\mbox{\textbf{$-0.70$}}\end{array}$ & $\def\arraystretch{0.6}-0.59\!\begin{array}{l}\mbox{\textbf{$+0.63$}}\\\mbox{\textbf{$-0.54$}}\end{array}$ & $\def\arraystretch{0.6}-0.12\!\begin{array}{l}\mbox{\textbf{$+0.41$}}\\\mbox{\textbf{$-0.35$}}\end{array}$  \\
      3C 435 & 0.4710 & FR II & $\def\arraystretch{0.6}+0.06\!\begin{array}{l}\mbox{\textbf{$+0.31$}}\\\mbox{\textbf{$-0.29$}}\end{array}$ & $\def\arraystretch{0.6}+0.70\!\begin{array}{l}\mbox{\textbf{$+0.37$}}\\\mbox{\textbf{$-0.40$}}\end{array}$ & $\def\arraystretch{0.6}-0.24\!\begin{array}{l}\mbox{\textbf{$+0.26$}}\\\mbox{\textbf{$-0.25$}}\end{array}$ & $\def\arraystretch{0.6}+0.26\!\begin{array}{l}\mbox{\textbf{$+0.18$}}\\\mbox{\textbf{$-0.18$}}\end{array}$  \\
      3C 436 & 0.2145 & FR II & $\def\arraystretch{0.6}+0.24\!\begin{array}{l}\mbox{\textbf{$+0.11$}}\\\mbox{\textbf{$-0.11$}}\end{array}$ & $\def\arraystretch{0.6}+0.16\!\begin{array}{l}\mbox{\textbf{$+0.11$}}\\\mbox{\textbf{$-0.11$}}\end{array}$ & $\def\arraystretch{0.6}+0.09\!\begin{array}{l}\mbox{\textbf{$+0.06$}}\\\mbox{\textbf{$-0.06$}}\end{array}$ & $\def\arraystretch{0.6}+0.01\!\begin{array}{l}\mbox{\textbf{$+0.05$}}\\\mbox{\textbf{$-0.05$}}\end{array}$  \\
      3C 438 & 0.2900 & FR II & $\def\arraystretch{0.6}-0.15\!\begin{array}{l}\mbox{\textbf{$+0.04$}}\\\mbox{\textbf{$-0.04$}}\end{array}$ & $-0.12\mbox{\textbf{$\pm0.05$}}^\dagger$ & $+0.06\mbox{\textbf{$\pm0.05$}}^\dagger$ & $-0.10\mbox{\textbf{$\pm0.05$}}^\dagger$  \\
      3C 441 & 0.7080 & FR II & $\def\arraystretch{0.6}+0.38\!\begin{array}{l}\mbox{\textbf{$+0.47$}}\\\mbox{\textbf{$-0.47$}}\end{array}$ & $\def\arraystretch{0.6}+0.54\!\begin{array}{l}\mbox{\textbf{$+0.45$}}\\\mbox{\textbf{$-0.49$}}\end{array}$ & $\def\arraystretch{0.6}-0.77\!\begin{array}{l}\mbox{\textbf{$+0.32$}}\\\mbox{\textbf{$-0.29$}}\end{array}$ & $\def\arraystretch{0.6}+0.63\!\begin{array}{l}\mbox{\textbf{$+0.23$}}\\\mbox{\textbf{$-0.23$}}\end{array}$  \\
      3C 455 & 0.5430 & FR II & $\def\arraystretch{0.6}+0.46\!\begin{array}{l}\mbox{\textbf{$+0.27$}}\\\mbox{\textbf{$-0.27$}}\end{array}$ & $\def\arraystretch{0.6}+0.22\!\begin{array}{l}\mbox{\textbf{$+0.36$}}\\\mbox{\textbf{$-0.36$}}\end{array}$ & $\def\arraystretch{0.6}+0.21\!\begin{array}{l}\mbox{\textbf{$+0.25$}}\\\mbox{\textbf{$-0.23$}}\end{array}$ & $\def\arraystretch{0.6}+0.23\!\begin{array}{l}\mbox{\textbf{$+0.17$}}\\\mbox{\textbf{$-0.16$}}\end{array}$ 
      \\
      \cline{1-7}
      \multicolumn{7}{l}{$^\dagger$Approximation Equation (\ref{eq:r_approx}) used for $S_{\rm{ON}}+S_{\rm{OFF}}>2000.$
      }
\end{longtable}

\section{Science aperture properties, X-ray and radio observation data.}
\label{ap:science_circles}
Table \ref{tab:raw_quadrants} shows the properties of the science aperture and the X-ray and radio data used for each source. Table \ref{tb:on_off_b_counts} shows the on-jet axis counts, off-jet axis counts, and the background level counts for the sources in our sample.

\startlongtable
\begin{deluxetable*}{llcccccc}
\centering
\tablecolumns{8} 
\setlength\tabcolsep{2.5pt}
\tabletypesize{\small}
\tablecaption{{Science aperture properties, X-ray and radio observation data.} \label{tab:raw_quadrants}}
\tablehead{ 
\colhead{Source}  & \colhead{Quadrant\tablenotemark{a}} & \colhead{Position\tablenotemark{b}} & \colhead{Quadrant \tablenotemark{c}} 
& \colhead{$Chandra$\tablenotemark{d}}  & \colhead{Exposure} 
& \colhead{Obs. date} & \colhead{Radio\tablenotemark{e}} \\
\colhead{} & \colhead{centroid} & \colhead{angle ($^\circ$)} & \colhead{radius (arcsec)} 
& \colhead{obs. ID} & \colhead{time (ks)} 
& \colhead{} & \colhead{obs freq. (GHz)}
}
\startdata 
3C 16 &  0:37:45.367,+13:20:09.43  & 30.92 & 54.36  & 13879  & 11.92 &    2012-10-25  & 1.50 \\
3C 19   &  0:40:55.056,+33:10:07.36  & 30.23 & 4.68 & 13880   & 11.90 &   2011-10-29  & 1.50 \\
3C 20  &  0:43:09.171,+52:03:36.25  & 105.53 & 30.03 & 9294   & 7.93 &   2007-12-31  & 8.40 \\
3C 22   &  0:50:56.220,+51:12:03.64  & 105.68 & 17.14 & 14994  & 9.35 &   2013-06-05   & 1.49 \\
3C 29   &  0:57:34.897,-1:23:27.29  & 163.69 & 90.28 & 12721  & 7.95 &   2011-5-23  & 1.50 \\
3C 34   &  1:10:18.540,+31:47:19.51  & 84.02 & 28.62 & 16046  & 11.92 &   2014-09-25  & 4.86 \\
3C 42   &  1:28:30.103,+29:03:01.16  & 133.29 & 18.68 & 13872 & 11.92 &   2012-02-26   & 1.50 \\
3C 46   &  1:35:28.506,+37:54:04.99  & 65.62 & 103.99 & 13881  & 11.92 &    2012-09-28  & 1.50 \\
3C 52   &  1:48:28.909,+53:32:28.04  & 16.60 & 59.54 & 9296  & 8.02 &   2008-03-26  & 1.58 \\
3C 54   &  1:55:30.285,+43:45:58.70  & 27.07 & 42.85 & 16049  & 11.92 &   2014-06-15  & 4.86 \\
3C 55   &  1:57:10.558,+28:51:39.49  & 98.14 & 47.92  & 16050  & 11.92 &   2014-06-15  & 4.87 \\
3C 79  &  3:10:00.102,+17:05:58.67  & 106.64 & 52.92  & 12723  & 7.69 &    2010-11-01  & 8.41 \\
3C 129.1 &  4:50:06.648,+45:03:06.07  & 35.58 & 57.75 &  2219   & 9.63 &    2001-01-09  & 4.74 \\
3C 133   &  5:02:58.470,+25:16:25.36  & 97.53 & 16.28  & 9300 &  8.03 &   2008-04-07  & 5.00 \\
3C 135   &  5:14:08.350,+0:56:32.89  & 75.83 & 83.52 & 9301  & 7.93 &   2008-01-11  & 8.40 \\
3C 166   &  6:45:24.098,+21:21:51.14  & 180.08 & 28.44 & 12727   & 7.95 &    2010-11-30  & 1.50 \\
3C 169.1  &  6:51:14.801,+45:09:28.39  & 141.70 & 33.84 & 16056  & 11.92 &   2014-08-17  & 1.50 \\
3C 171  &  6:55:14.746,+54:08:57.71  & 100.43 & 25.25 & 10303  & 59.46 &   2009-01-08  & 1.50 \\
3C 175.1  &  7:14:04.709,+14:36:22.80  & 71.51 & 7.20 & 15000  & 9.94 &   2013-02-10  & 5.00 \\
3C 184.1   &  7:43:01.368,+80:26:25.86  & 157.78 & 111.60  & 9305  & 8.02 &   2008-03-27  & 1.50 \\
3C 192  &  8:05:34.999,+24:09:50.30  & 124.38 & 112.10  & 9270  & 10.02 &   2007-12-18  & 8.40 \\
3C 215  &  9:06:31.877,+16:46:11.89  & 149.10 & 34.84  & 3054  & 33.80 &   2003-01-02  & 4.86 \\
3C 223.1  &  9:41:24.029,+39:44:42.04  & 14.21 & 43.20  & 9308 & 7.93 &   2008-01-16  & 8.40 \\
3C 234  &  10:01:49.519,+28:47:08.64  & 61.58 & 69.71  & 12732  & 7.95 &   2011-01-19  & 1.50 \\
3C 244.1  &  10:33:34.022,+58:14:35.36  & 166.90 & 32.99  & 13882  & 11.92 &   2013-01-16   & 1.50 \\
3C 268.2   &  12:00:58.754,+31:33:21.34  & 22.02 & 56.50  & 13876   & 11.58 &   2012-07-07  & 1.50 \\
3C 272   &  12:24:28.471,+42:06:35.89  & 22.25 & 48.71  & 16061   & 11.92 &   2015-03-01  & 1.50 \\
3C 277  &  12:51:42.002,+50:34:24.99  & 73.82 & 94.81  & 16062 & 11.92 &   2015-03-03   & 1.50 \\
3C 284   &  13:11:04.632,+27:28:07.31  & 99.95 & 133.24  & 12735   & 7.95 &   2010-11-17  & 1.50 \\
3C 285   &  13:21:17.866,+42:35:15.17  & 76.23 & 110.24  & 6911  & 39.62 &   2006-06-18  & 1.42 \\
3C 287.1  &  13:32:53.261,+2:00:45.68  & 92.56 & 79.20  & 9309  & 8.02 &   2008-03-23  & 1.50 \\
3C 289  &  13:45:26.266,+49:46:32.79  & 109.53 & 6.87  & 15007  & 9.70 &   2013-07-28  & 8.44 \\
3C 292  &  13:50:41.873,+64:29:35.88  & 164.74 & 88.25  & 16065, 17488  & 3.97, 7.97 &  2014-09-12, 2014-11-21  & 8.46 \\
3C 293.1  &  13:54:40.999,+16:14:50.00  & 45.25 & 47.47  & 16066  & 11.92 &   2014-06-05  & 1.44 \\
3C 300  &  14:22:59.861,+19:35:36.94  & 137.48 & 72.71  & 9311 & 7.94 &   2008-03-21  & 8.33 \\
3C 306.1  &  14:55:01.409,-4:21:00.02  & 1.13 & 64.21  & 13885  & 11.92 &   2012-09-06   & 1.40 \\
3C 315  &  15:13:40.068,+26:07:30.25  & 158.98 & 100.66  & 9315  & 7.67 &   2010-12-10  & 1.40 \\
3C 319  &  15:24:05.028,+54:28:04.72  & 44.13 & 66.43 & 12737  & 8.44 &   2010-10-25  & 8.33 \\
3C 323.1  &  15:47:43.518,+20:52:16.70  & 20.72 & 51.70  & 9314  & 7.93 &   2008-06-01  & 4.86 \\
3C 327.1  &  16:04:45.408,+1:17:50.13  & 109.52 & 14.03  & 13887  & 11.11 &   2012-05-05  & 1.43 \\
3C 332   &  16:17:42.552,+32:22:34.35  & 16.70 & 57.16  & 9315  & 7.93 &   2007-12-10  & 4.86 \\
3C 336   &  16:24:39.084,+23:45:12.37  & 35.07 & 16.44  & 15008  & 2.00 &   2013-03-03  & 8.44 \\
3C 340   &  16:29:36.588,+23:20:12.90  & 86.16 & 29.59  & 15010  & 9.95 &   2013-10-20  & 8.44 \\
3C 341   &  16:28:03.994,+27:41:39.30  & 45.23 & 46.80  & 13888  & 11.64 &   2011-11-14  & 1.45 \\
3C 352   &  17:10:44.126,+46:01:28.75  & 161.15 & 7.79  & 15013   & 9.95 &   2013-10-10  & 8.46 \\
3C 357   &  17:28:20.105,+31:46:02.64  & 100.61 & 57.06  & 12738  & 7.95 &   2010-10-31  & 4.89 \\
3C 381   &  18:33:46.313,+47:27:02.65  & 3.20 & 43.19  & 9317  & 8.06 &    2008-02-21  & 1.42 \\
3C 382   &  18:35:03.389,+32:41:46.83  & 49.93 & 98.16  & 4910, 6151  & 54.15, 63.87 &   2004-10-27, 2004-10-30 & 8.46 \\
3C388   &  18:44:02.378,+45:33:29.95  & 61.60 & 31.53  & 4756, 5295  & 7.37, 30.71 &    2004-01-09, 2004-01-29 & 1.39 \\
3C 401   &  19:40:25.015,+60:41:36.00  & 25.50 & 14.19  & 3083, 4370  & 22.66, 24.85&   2002-09-20, 2002-09-21 & 8.44 \\
3C 411   &  20:22:08.455,+10:01:11.58  & 114.29 & 18.23  & 13889  & 11.92 &   2012-08-08  & 1.50 \\
3C 424   &  20:48:12.101,+7:01:17.13  & 158.05 & 22.68  & 12743  & 7.95 &   2011-04-15  & 8.33 \\
3C 427.1  &  21:04:06.924,+76:33:10.33  & 140.39 & 15.85  & 2194  & 39.45 &   2002-01-27  & 5.00 \\
3C 432  &  21:22:46.333,+17:04:37.73  & 132.89 & 9.63  & 5624  & 19.78 &   2005-01-07  & 1.48 \\
3C 434   & 21:23:16.241,+15:48:05.75  & 77.89 & 11.46  & 13878   & 11.92 &   2012-08-15  & 8.44 \\
3C 435  &  21:29:06.115,+7:32:54.96  & 54.38 & 28.69  & 13890  & 11.92 &   2012-08-14  & 1.50 \\
3C 436  &  21:44:11.717,+28:10:19.47  & 174.45 & 62.99  & 9318, 12745  & 8.04, 7.95 &   2008-01-08, 2011-05-27 & 1.50 \\
3C 438  &  21:55:52.255,+38:00:28.20  & 126.00 & 12.93 & 12879, 13218  & 72.04, 47.44 &   2011-01-30, 2011-01-28 & 1.50 \\
3C 441  &  22:06:04.916,+29:29:19.98  & 147.48 & 24.46  & 14993, 15656 & 3.03, 6.98 &   2013-05-24, 2013-06-26 & 1.45 \\
3C 455  &  22:55:03.888,+13:13:34.13  & 57.63 & 3.50  & 15014  & 9.95 &   2013-08-13  & 8.44 \\
\enddata 
\tablenotetext{a}{Quadrant centroid position in RA and DEC (J2000).}
\tablenotetext{b}{Position angle of the radio jet measured East of North in degrees.}
\tablenotetext{c}{Quadrant radius in arcseconds.}
\tablenotetext{d}{$Chandra$ observation ID of the X-ray data.}
\tablenotetext{e}{Frequency of the VLA radio observation.}
\end{deluxetable*} 

\startlongtable
\begin{deluxetable*}{l|ccc|ccc|ccc|ccc}
   \centering
   \tablecolumns{13}
   \tablewidth{700pt}
   \tabletypesize{\small}
   \tablecaption{Science aperture ON, OFF, and $B$ counts for the soft, medium, hard, and broad bands.\label{tb:on_off_b_counts}}
   \tablehead{
      \colhead{Source} &
      \multicolumn{3}{c}{Soft ($0.5 - 1.2$ keV)} &
      \multicolumn{3}{c}{Medium ($1.2 - 2.0$ keV)} &
      \multicolumn{3}{c}{Hard ($2.0 - 7.0$ keV)} &
      \multicolumn{3}{c}{Broad ($0.5 - 7.0$ keV)} \\
      \colhead{} &
      \colhead{ON} &
      \colhead{OFF} &
      \colhead{$B$} &
      \colhead{ON} &
      \colhead{OFF} &
      \colhead{$B$} &
      \colhead{ON} &
      \colhead{OFF} &
      \colhead{$B$} &
      \colhead{ON} &
      \colhead{OFF} &
      \colhead{$B$}
   }
   \startdata
      3C 16 & 31 & 13 & 28 & 20 & 22 & 31 & 52 & 57 & 101 & 108 & 88 & 160 \\
      3C 19 & 11 & 14 & 1 & 12 & 8 & 1 & 3 & 4 & 1 & 26 & 24 & 3 \\
      3C 20 & 13 & 14 & 8 & 11 & 7 & 7 & 26 & 25 & 27 & 51 & 48 & 42 \\
      3C 22 & 10 & 2 & 2 & 9 & 3 & 3 & 10 & 15 & 7 & 33 & 15 & 13 \\
      3C 29 & 97 & 49 & 80 & 68 & 71 & 105 & 111 & 111 & 207 & 277 & 231 & 391 \\
      3C 34 & 14 & 8 & 10 & 14 & 17 & 8 & 15 & 25 & 20 & 58 & 40 & 39 \\
      3C 42 & 9 & 1 & 6 & 2 & 5 & 5 & 12 & 13 & 16 & 25 & 19 & 27 \\
      3C 46 & 83 & 85 & 105 & 64 & 77 & 114 & 218 & 204 & 362 & 353 & 378 & 581 \\
      3C 52 & 42 & 26 & 30 & 57 & 51 & 40 & 73 & 78 & 104 & 177 & 150 & 173 \\
      3C 54 & 10 & 9 & 13 & 19 & 8 & 22 & 39 & 48 & 56 & 78 & 55 & 90 \\
      3C 55 & 27 & 8 & 19 & 25 & 8 & 23 & 35 & 42 & 68 & 94 & 53 & 110 \\
      3C 79 & 27 & 22 & 23 & 25 & 23 & 22 & 51 & 57 & 77 & 111 & 96 & 122 \\
      3C 129.1 & 70 & 69 & 65 & 254 & 251 & 248 & 381 & 441 & 353 & 766 & 702 & 666 \\
      3C 133 & 6 & 5 & 2 & 8 & 9 & 4 & 12 & 22 & 9 & 40 & 28 & 14 \\
      3C 135 & 52 & 51 & 86 & 49 & 42 & 79 & 144 & 131 & 298 & 231 & 236 & 474 \\
      3C 166 & 21 & 11 & 5 & 25 & 13 & 9 & 16 & 22 & 26 & 67 & 40 & 41 \\
      3C 169.1 & 17 & 6 & 7 & 13 & 5 & 12 & 22 & 26 & 36 & 56 & 33 & 55 \\
      3C 171 & 65 & 56 & 44 & 51 & 38 & 58 & 123 & 110 & 159 & 228 & 219 & 261 \\
      3C 175.1 & 2 & 4 & 0 & 7 & 4 & 1 & 6 & 6 & 2 & 15 & 13 & 3 \\
      3C 184.1 & 81 & 65 & 123 & 77 & 61 & 130 & 233 & 225 & 344 & 384 & 360 & 596 \\
      3C 192 & 154 & 103 & 123 & 111 & 93 & 164 & 277 & 303 & 552 & 568 & 472 & 839 \\
      3C 215 & 228 & 221 & 36 & 158 & 121 & 43 & 161 & 199 & 128 & 581 & 503 & 207 \\
      3C 223.1 & 14 & 11 & 19 & 19 & 12 & 16 & 39 & 45 & 58 & 79 & 60 & 93 \\
      3C 234 & 54 & 47 & 37 & 39 & 33 & 45 & 95 & 107 & 140 & 203 & 173 & 223 \\
      3C 244.1 & 26 & 17 & 15 & 22 & 17 & 11 & 25 & 37 & 36 & 83 & 60 & 62 \\
      3C 268.2 & 28 & 22 & 35 & 37 & 25 & 38 & 62 & 57 & 116 & 121 & 109 & 189 \\
      3C 272 & 15 & 6 & 21 & 18 & 15 & 15 & 38 & 44 & 66 & 77 & 59 & 103 \\
      3C 277 & 66 & 52 & 85 & 66 & 61 & 77 & 135 & 163 & 243 & 295 & 248 & 404 \\
      3C 284 & 108 & 81 & 145 & 87 & 104 & 169 & 241 & 274 & 497 & 484 & 410 & 811 \\
      3C 285 & 482 & 486 & 504 & 318 & 337 & 523 & 859 & 847 & 1661 & 1650 & 1686 & 2689 \\
      3C 287.1 & 124 & 88 & 58 & 102 & 64 & 63 & 143 & 162 & 190 & 391 & 301 & 311 \\
      3C 289 & 2 & 1 & 0 & 6 & 4 & 1 & 5 & 9 & 1 & 17 & 9 & 2 \\
      3C 292 & 83 & 57 & 109 & 90 & 55 & 93 & 182 & 181 & 311 & 357 & 298 & 514 \\
      3C 293.1 & 28 & 19 & 34 & 30 & 14 & 23 & 42 & 38 & 72 & 97 & 75 & 129 \\
      3C 300 & 50 & 35 & 67 & 43 & 25 & 64 & 87 & 86 & 170 & 180 & 148 & 301 \\
      3C 306.1 & 38 & 35 & 47 & 34 & 23 & 42 & 46 & 89 & 116 & 161 & 104 & 205 \\
      3C 315 & 75 & 65 & 78 & 68 & 63 & 115 & 144 & 181 & 310 & 324 & 273 & 503 \\
      3C 319 & 40 & 35 & 41 & 37 & 24 & 42 & 63 & 72 & 124 & 149 & 122 & 208 \\
      3C 323.1 & 175 & 168 & 31 & 112 & 129 & 28 & 110 & 123 & 75 & 424 & 417 & 134 \\
      3C 327.1 & 53 & 54 & 3 & 55 & 34 & 5 & 36 & 38 & 7 & 145 & 129 & 15 \\
      3C 332 & 90 & 71 & 31 & 129 & 77 & 38 & 234 & 215 & 96 & 433 & 382 & 164 \\
      3C 336 & 0 & 5 & 0 & 7 & 0 & 0 & 2 & 5 & 4 & 12 & 7 & 5 \\
      3C 340 & 9 & 5 & 7 & 14 & 4 & 9 & 12 & 24 & 22 & 47 & 22 & 38 \\
      3C 341 & 19 & 20 & 22 & 19 & 10 & 28 & 35 & 49 & 81 & 88 & 66 & 131 \\
      3C 352 & 6 & 2 & 1 & 8 & 3 & 0 & 9 & 2 & 3 & 19 & 15 & 4 \\
      3C 357 & 35 & 26 & 30 & 39 & 17 & 35 & 54 & 57 & 103 & 128 & 96 & 168 \\
      3C 381 & 23 & 11 & 24 & 15 & 8 & 21 & 32 & 36 & 60 & 75 & 52 & 104 \\
      3C 382 & 4280 & 4027 & 4388 & 5270 & 4389 & 6128 & 4742 & 5158 & 5319 & 8014 & 7234 & 8360 \\
      3C 388 & 743 & 765 & 31 & 834 & 772 & 33 & 497 & 530 & 58 & 2111 & 2030 & 122 \\
      3C 401 & 374 & 478 & 14 & 244 & 265 & 7 & 174 & 150 & 25 & 777 & 919 & 47 \\
      3C 411 & 37 & 32 & 2 & 43 & 45 & 4 & 43 & 45 & 12 & 130 & 119 & 19 \\
      3C 424 & 13 & 14 & 5 & 14 & 8 & 4 & 13 & 14 & 12 & 41 & 30 & 21 \\
      3C 427.1 & 50 & 44 & 9 & 49 & 34 & 8 & 29 & 45 & 22 & 144 & 106 & 39 \\
      3C 432 & 23 & 13 & 2 & 16 & 14 & 2 & 17 & 17 & 5 & 62 & 43 & 8 \\
      3C 434 & 2 & 7 & 3 & 3 & 2 & 1 & 3 & 5 & 7 & 10 & 11 & 11 \\
      3C 435 & 17 & 16 & 11 & 13 & 7 & 9 & 22 & 27 & 23 & 55 & 44 & 43 \\
      3C 436 & 164 & 134 & 221 & 154 & 135 & 236 & 574 & 532 & 972 & 848 & 844 & 1429 \\
      3C 438 & 759 & 882 & 18 & 1236 & 1393 & 31 & 1328 & 1261 & 74 & 3264 & 3610 & 123 \\
      3C 441 & 7 & 5 & 5 & 8 & 5 & 5 & 10 & 20 & 21 & 35 & 20 & 31 \\
      3C 455 & 16 & 10 & 0 & 10 & 8 & 0 & 21 & 17 & 1 & 44 & 35 & 2 \\
   \enddata
\end{deluxetable*}

\bibliographystyle{aasjournal}
\bibliography{references}

\end{document}